\documentclass[reprint,amsmath,amssymb]{revtex4-1}
\usepackage{graphicx,grffile}
\usepackage{dcolumn}
\usepackage{bm}
\usepackage{subfig}
\usepackage{caption}
\usepackage{xcolor}
\usepackage{float}

\begin{document}

\title{Environment induced Symmetry Breaking of the Oscillation-Death State}
\author{Sudhanshu Shekhar Chaurasia$^1$}

\author{Manish Yadav$^1$}
 
\author{Sudeshna Sinha$^1$}

\affiliation{
        \vspace{1 mm}$^1$Indian Institute of Science Education and Research (IISER) Mohali,\\
        Knowledge City, SAS Nagar, Sector 81, Manauli PO 140 306, Punjab, India
    }

\begin{abstract}
We investigate the impact of a common external system, which we call a common environment, on the Oscillator Death (OD) states of a group of Stuart-Landau oscillators. The group of oscillators yield a completely symmetric OD state when uncoupled to the external system, i.e. the two OD states occur with equal probability. However, remarkably, when coupled to a common external system this symmetry is significantly broken. For exponentially decaying external systems, the symmetry breaking is very pronounced for low environmental damping and strong oscillator-environment coupling. This is evident through the sharp transition from the symmetric to asymmetric state occurring at a critical oscillator-environment coupling strength and environmental damping rate. Further, we consider time-varying connections to the common external environment, with a fraction of oscillator-environment links switching on and off. Interestingly, we find that the asymmetry induced by environmental coupling decreases as a power law with increase in fraction of such on-off connections. The suggests that blinking oscillator-environment links can restore the symmetry of the OD state. Lastly, we demonstrate the generality of our results for a constant external drive, and find marked breaking of symmetry in the OD states there as well. When the constant environmental drive is large, the asymmetry in the OD states is very large, and the transition between the symmetric and asymmetric state with increasing oscillator-environment coupling is very sharp. So our results demonstrate an environmental coupling-induced mechanism for the prevalence of certain OD states in a system of oscillators, and suggests an underlying process for obtaining certain states preferentially in ensembles of oscillators with environment-mediated coupling.
\end{abstract}

\maketitle


\section{Introduction}
  Complex systems has been a very active area of research 
over the past few decades, initiated by the discovery that even systems with low degrees of freedom can show a wide range of dynamical patterns.
For instance, two or more oscillators, when coupled to each other can show completely synchronized oscillations, in-phase or anti-phase synchronized oscillations, oscillation quenching to homogeneous steady states or inhomogeneous steady states, with transitions between different dynamical behaviours obtained by parameter tuning. 

In general, oscillation quenching is categorized into homogeneous steady state (HSS) or amplitude death (AD) and inhomogeneous steady state (IHSS) or Oscillation death (OD) \cite{AD_OD}. AD refers to the situation where the coupled oscillator systems, under oscillation quenching, evolve to the same fixed point. This type of quenching is relevant in laser systems \cite{laser1,laser2,laser3}, and is important in situations involving stabilization to a particular fixed point. A lot of mechanisms leading to amplitude death have been found, such as time-delay in the coupling \cite{AD_time_delay1,AD_time_delay2}, coupling 
via conjugate variables \cite{conjugate}, introduction of large variance of frequencies \cite{AD_freq} and coupling to a dissimilar external oscillator \cite{chaurasia}. 
On the other hand, oscillation quenching can give rise to oscillation death, a phenomenon that is completely different from AD. Here the oscillators split into two sub-groups, around an unstable fixed point via pitchfork bifurcations, generating a set of stable fixed points. Oscillation death is very relevant to biological systems, as this oscillation quenching mechanism can lead to the emergence of inhomogeneity in homogeneous medium. So, for instance, OD has been interpreted as a mechanism for cellular differentiation \cite{cell_diff1,cell_diff2}. Thus a lot of research effort has centered around transitions from AD to OD \cite{AD_OD_transtion1,AD_OD_transtion2}, and mechanisms that steer the dynamics to the OD state have been investigated. For example, OD can be achieved via parametric modulation in coupled non-autonomous system \cite{paramertic_modulation}, parameter mismatch (i.e. detuning of parameters) in coupled oscillators \cite{parameter_detuning1,parameter_detuning2} and the introduction of local repulsive links in diffusively coupled oscillators \cite{repulsive_link}. In a complementary direction, some studies have also shown how OD states are eliminated when gradient coupling is introduced in delay induced OD \cite{eliminate_OD}.

Our work here focuses on oscillation quenching mechanisms that give rise to inhomogeneous steady states. Our test-bed will be a group of oscillators, coupled to a common external system, which is dynamically very distinct from the oscillators. This common external system provides a common ``environment'' and allows a group of oscillators to be indirectly coupled via an external common medium. When uncoupled, the oscillators have equal probability to go to either of the OD states. However, we will show that this system displays {\em symmetry breaking} when coupled. That is, a specific oscillator death state is preferentially achieved. This state selection leads to asymmetric distribution of OD states in the ensemble of oscillators, suggesting a natural mechanism that allows the emergence of a favoured set of fixed points. Further we will explore the effect of the oscillator group connecting to the environment through links that switch on and off. We will demonstrate that blinking oscillator-environment connections will remarkably work to towards partial restoration of the symmetry of the oscillator death states, though the presence of some blinking connections reduces the symmetry of the dynamical equations.


\section{Coupled oscillators}
    Complex systems often undergo Hopf bifurcations and sufficiently close to such a bifurcation point, the variables which have slower time-scales can be eliminated. This leaves us with a couple of simple first order ordinary differential equations, popularly known as the Stuart-Landau system \cite{kuramoto}. In this work we  consider a group of $N$ globally coupled Stuart-Landau oscillators, with $\omega$ being the angular frequency of oscillator. Specifically, the oscillators are coupled via the mean field $\bar{x}$ of the $x$-variable, with $\varepsilon_\text{intra}$ reflecting the strength of intra-group coupling. Now, this oscillator group also couples to an external system, which we call the {\em environment}, denoted by $u$. The environment exponentially decays to zero, with decay constant $k$, when uncoupled from the oscillator group. However when coupled to the oscillators, the environment provides an input to the oscillators, as well as receives a feedback proportional to the mean field $\bar{y}$ of the $y$-variables of the oscillators.
The strength of this feedback from the external system is given by the coupling strength $\varepsilon_{\text{ext}}$. So the complete dynamics of the group of oscillators, along with the external environment, is then given by the following evolution equations:

    \begin{eqnarray}
        \label{eqn_system}
        \dot{x}_i &=& (1-x_i^2-y_i^2)x_i-\omega y_i + \varepsilon_\text{intra}(q\bar{x}-x_i) \nonumber \\
        \dot{y}_i &=& (1-x_i^2-y_i^2)y_i+\omega x_i + \varepsilon_\text{ext} u\\
        \dot{u} &=& -k u+ \bar{y} \nonumber
    \end{eqnarray}

    where $\bar{x}=\frac{1}{N}\sum_{i=1}^N x_i$    and    $\bar{y}=\frac{1}{N}\sum_{i=1}^N \varepsilon_\text{ext} y_i$. 
    
    So the common external environment provides an {\em indirect coupling} conjoining the different oscillators in the group. The idea is rooted in phenomena where a common medium influences oscillators, such as in the population of yeast cells \cite{medium_yeast}, where acetaldehyde is used as a common medium, or in mechanical oscillators in a fluid environment \cite{mech_fluid}. Studies on the effect of an external environment on coupled Stuart-Landau oscillators has revealed phenomena such as the revival of oscillations in a group of oscillators at steady state by coupling 
to an oscillating group via a common environment \cite{yadav2018revival}, phase-flip transitions 
in a system of oscillators diffusively coupled to the environment \cite{sharma2016phase} and co-existence of in-phase oscillations and oscillation death in environmentally coupled oscillators \cite{verma2018co}.

    In this work we will first explore in Section III the symmetry-breaking effect of the common external environment on the oscillatory patterns. Further, we will explore the spatiotemporal effects of the time variation of the oscillator-environment links in Section IV. Lastly in Section V, we will demonstrate  that a constant common environment also leads to pronounced symmetry breaking in the Oscillator Death states, suggesting the generality of our central result.
    

\section{Symmetry Breaking in the Oscillator Death States}

\begin{figure}[H]
\centering
\includegraphics[width=1.0\linewidth]{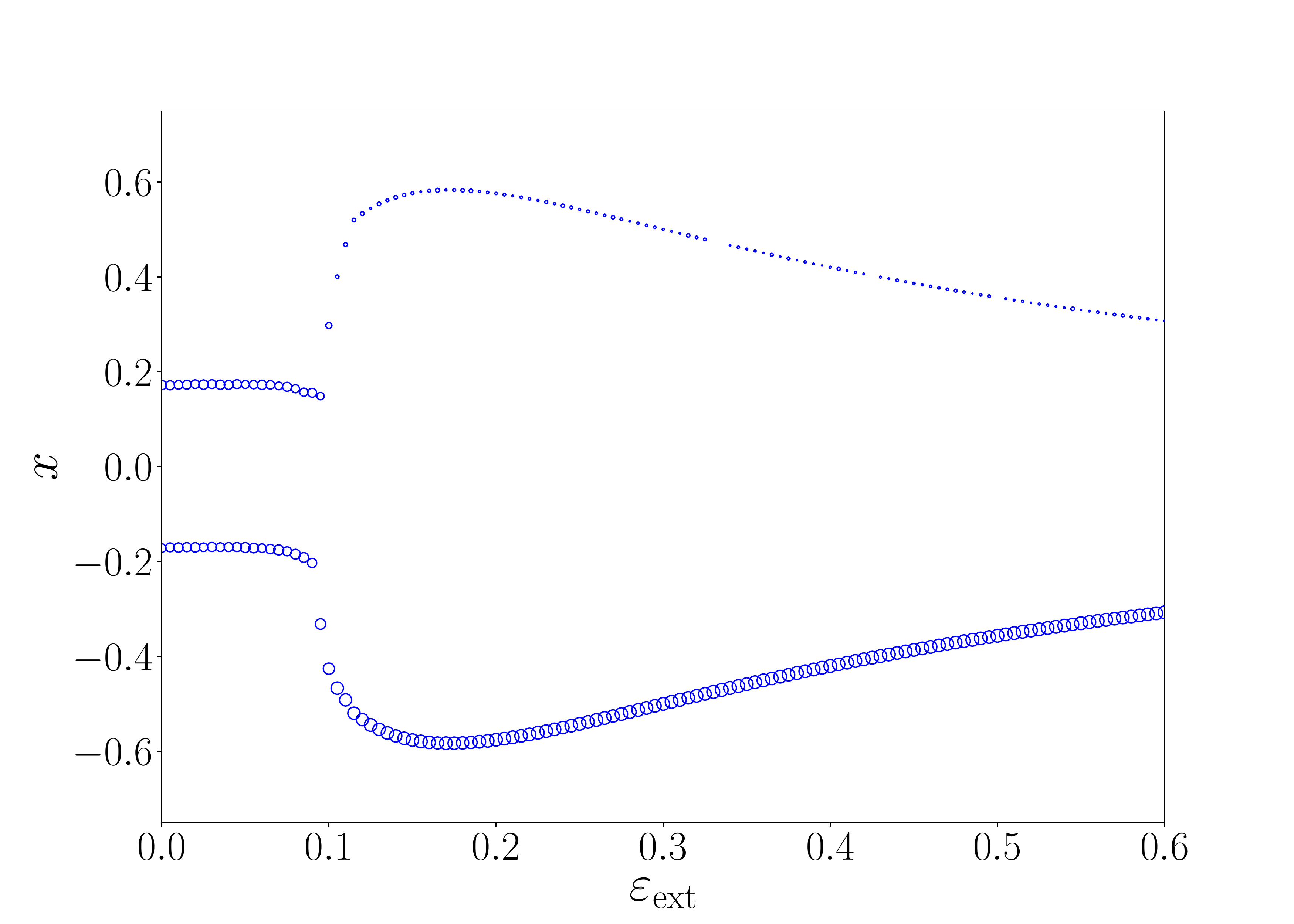}
\caption{Bifurcation diagram of $x$ of one of the Stuart-Landau oscillator in a group, with respect to the coupling strength $\varepsilon_\text{ext}$ of the group with the environment (cf. Eqn.~\ref{eqn_system}). The diagram displays the superposition of the system evolving from a large range of random initial states, with $x_{i}, y_{i} \in [-1,1]$  and the environmental variable $u \in [0,1]$. The size of the circle represent the probability of being in that state (positive or negative). Here we consider the Stuart-Landau oscillators with parameters $\omega$=$2.0$, $q = 0.4$ and $\varepsilon_\text{intra} = 6.0$ (namely in the Oscillator Death region when uncoupled to the environment). The environmental damping constant $k = 0.01$ and the system size $N=20$.}
\label{bifurcation}
\end{figure}

We first present the bifurcation sequence of the oscillators as a function of the oscillator-environment coupling strength $\varepsilon_\text{ext}$. The values of $\varepsilon_\text{intra}$ and $q$ are fixed at 6.0 and 0.4 so that uncoupled oscillators are in the oscillation death (OD) state in the absence of coupling to the environment. Here one of the oscillator death states has positive $x$ and negative $y$, and the other oscillator death state has and negative $x$ and positive $y$ (cf. Fig.~\ref{bifurcation}). We call the steady state solution with $x > 0$ the ``positive state'' and the steady state with $x < 0$  the ``negative state''. In the bifurcation diagram in Fig.~\ref{bifurcation}, the size of the symbols represent the probability of being in that state.

\begin{figure}[H]
    \centering
    \includegraphics[width=1.0\linewidth] {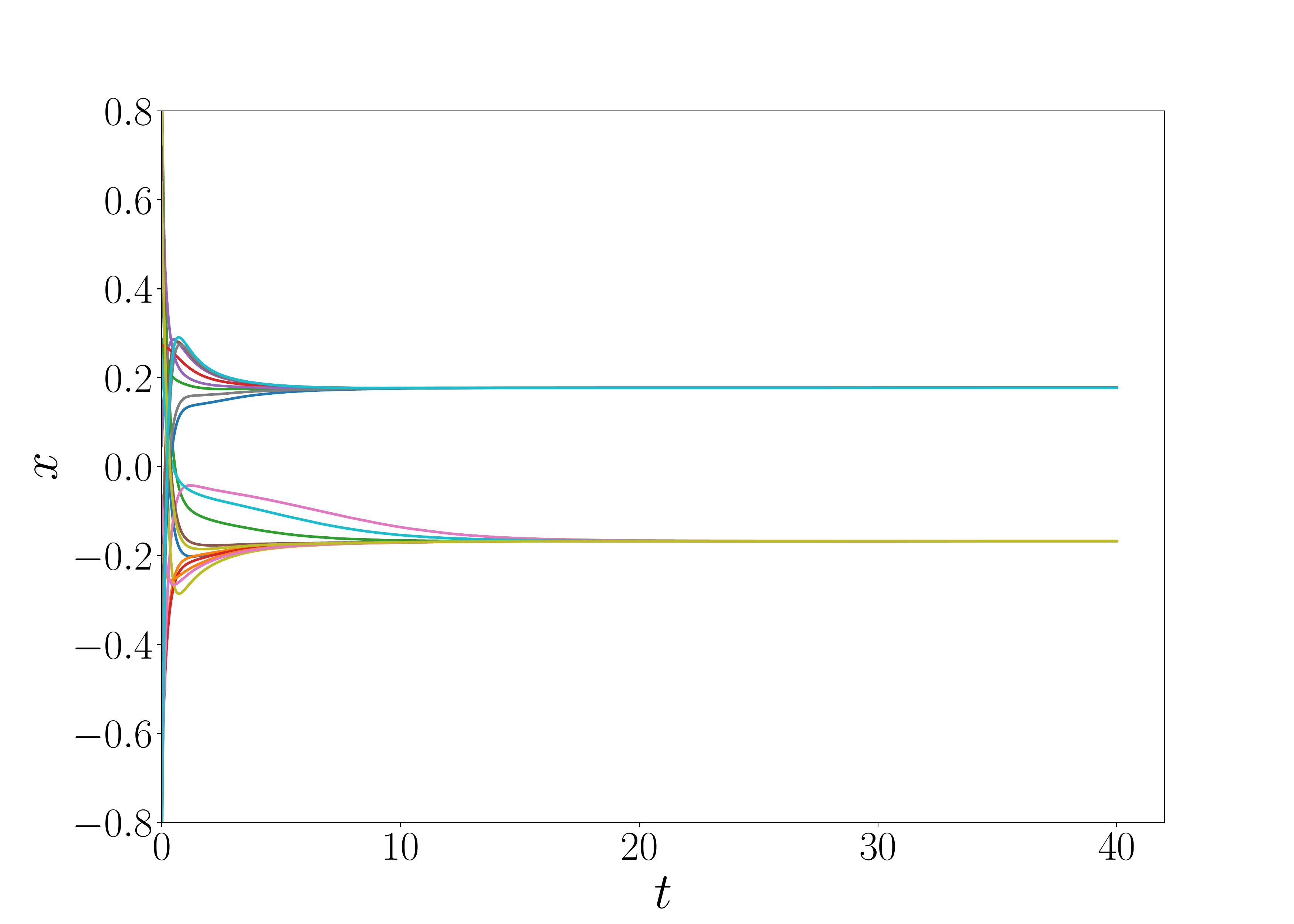}\\ 
     \hfill (a) \hfill $ $\\
    \includegraphics[width=1.0\linewidth] {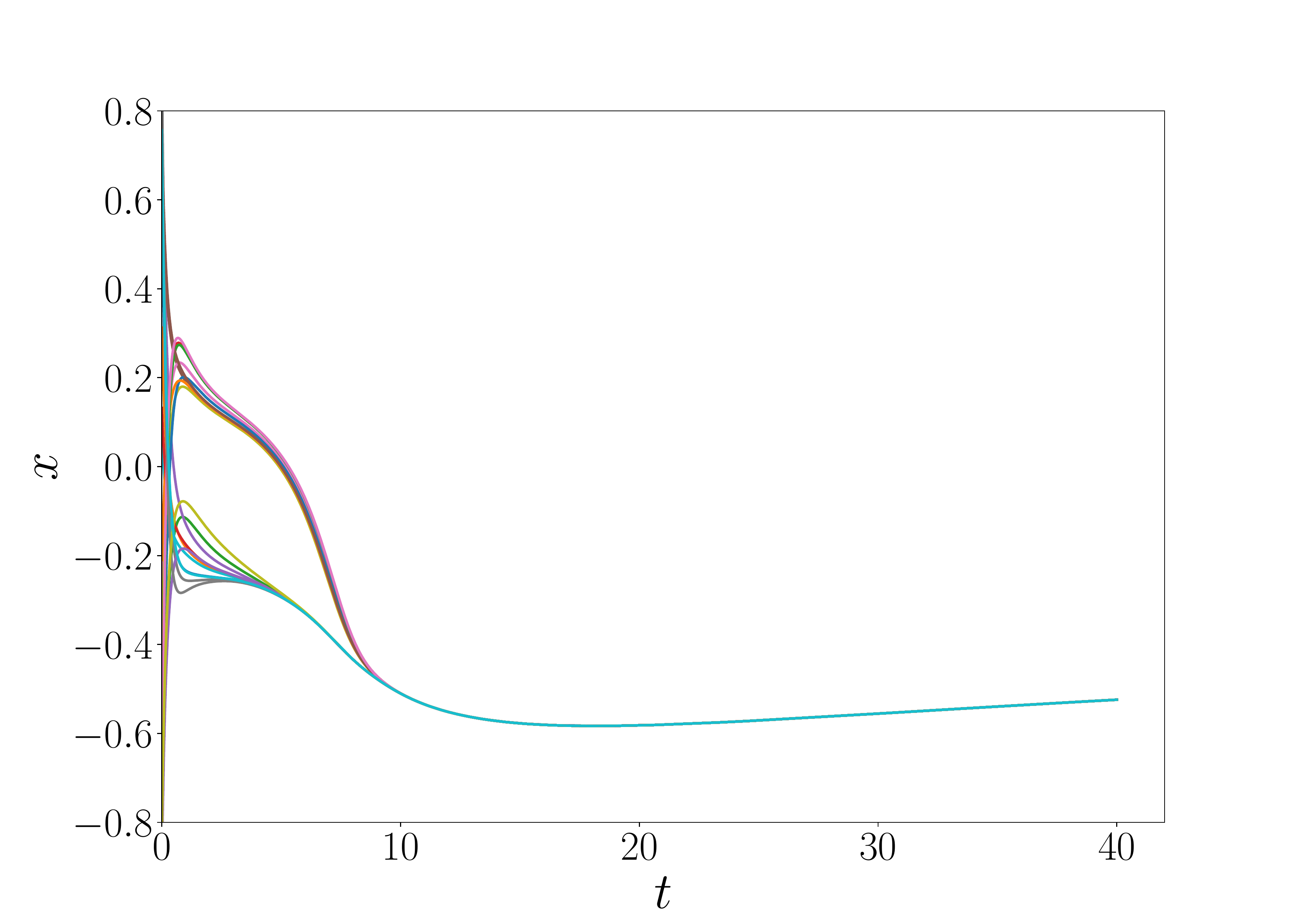}\\
    \hfill (b) \hfill $ $
	\caption{Time series of twenty oscillators in the group (shown in distinct colours), (a) in the absence of coupling to an external environment and (b) when the group is connected to the external environment with coupling strength $\varepsilon_\text{ext}=0.6$, and $k=0.01$.}
    \label{ts_time_series_no_blinking}
\end{figure}

In the absence of coupling to an external environment, the states of the group of oscillators are symmetrically distributed between the positive and negative states. That is, starting from generic random initial conditions, the group of oscillators will have equal probability to evolve to a positive state or a negative state. So one typically observes an equi-distribution of positive and negative oscillators in a group of Stuart-Landau oscillators in the oscillator death regime, when uncoupled to the environment. This is evident from the bifurcation diagram, which shows equal probability to be in either of the two OD states  at $\varepsilon_\text{ext} = 0$ (as reflected by symbols of the same size in the positive and negative states in the figure at $\varepsilon_\text{ext} = 0$). This behaviour is also clear from the time series of the oscillator group displayed in Fig.~\ref{ts_time_series_no_blinking}a.

Interestingly however, when the oscillator group is coupled to the external environment we observe {\em symmetry breaking in the Oscillator Death states}. Namely, the group of oscillators in the presence of the environment, preferentially go to one of the oscillator death state. So, typically we do not obtain an equal number of positive and negative states. Rather there is now a {\em pronounced prevalence of one of the Oscillator Death states}.  

This is evident from the bifurcation diagram, which shows unequal probability to be in the OD states, especially at large $\varepsilon_\text{ext}$. This is reflected by symbols of the different sizes in the positive and negative states in the figure at large $\varepsilon_\text{ext}$. This behaviour is also clear from the time series of the oscillator group displayed in Fig.~\ref{ts_time_series_no_blinking}b, which shows the oscillators preferentially evolving to one of the two OD solutions.



In order to gauge the global stability of an Oscillator Death state, say the positive state, we use the concept of Basin Stability. We choose a large number of random initial conditions, uniformly spread over phase space volume. For each initial state, we calculate the fraction $f^+$ of oscillators that evolve to the positive OD state. The average of $f^+$ over random initial conditions $\langle f^+ \rangle$ yields an estimate of the Basin Stability of the positive state, and indicates the probability of obtaining the positive oscillator death state in a group of oscillators starting from random initial conditions in the prescribed volume of phase space. The most symmetric distribution, namely half the oscillators in the positive state and the other half in the negative state, leads to a Basin Stability measure of $0.5$. {\em Deviations from $0.5$ indicate asymmetry in the distribution of oscillator death states}, with a prevalence of the positive or negative state. So the quantity $\langle f^+ \rangle$ serves as an order parameter for symmetry-breaking of the Oscillator Death states.



\begin{figure}[H]
    \centering
    \includegraphics[width=1.0\linewidth]
    {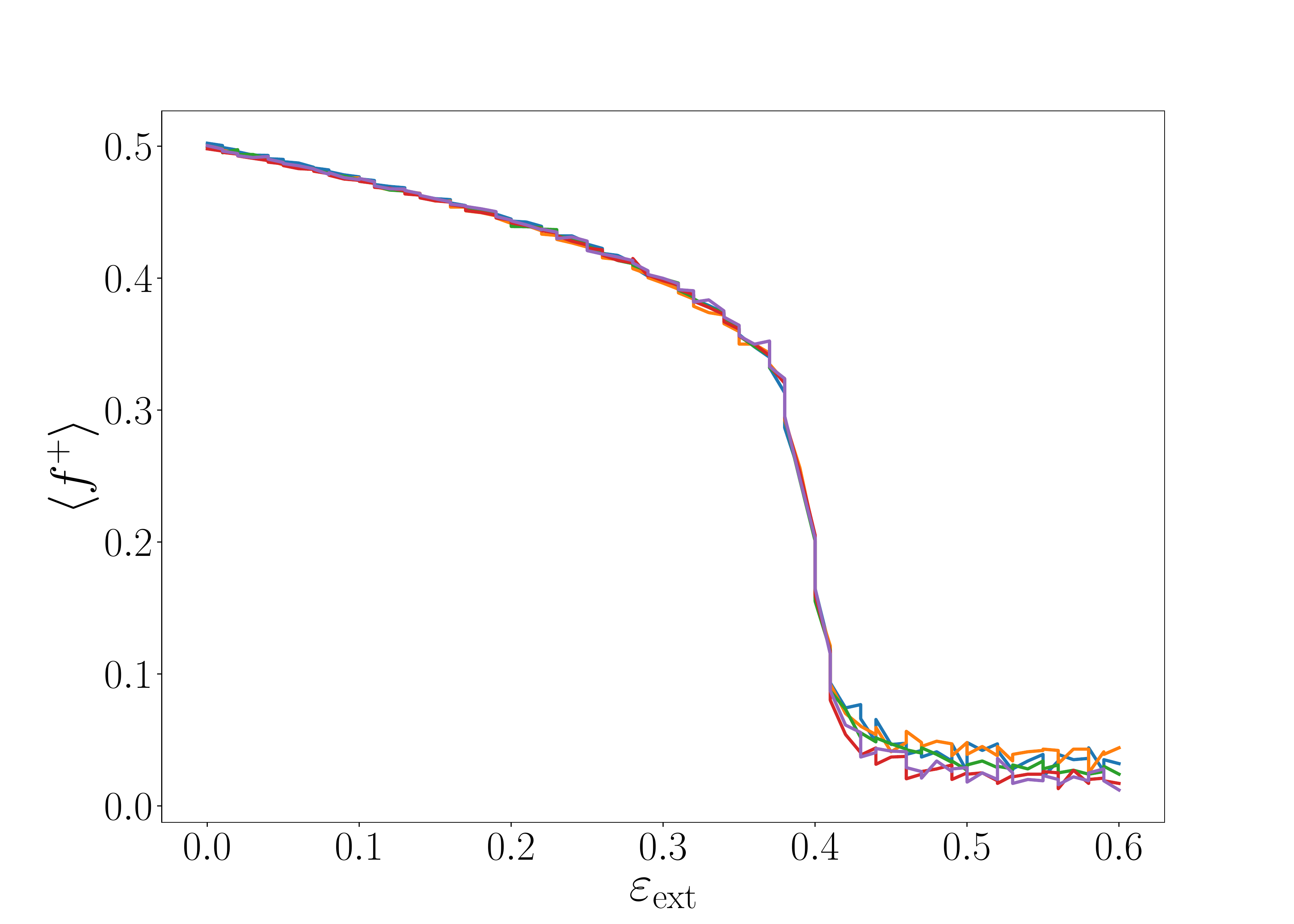}
    \caption{Basin Stability of the positive Oscillator Death state of coupled oscillators with $\varepsilon_\text{ext}$, where groups of oscillators of different sizes $N$=$40,60,80,100,120$ are shown in different colours. Here the damping constant of the external environment is $k=0.2$.}
    \label{bs_with_cp_for_different_N_without_blinking}
  \end{figure}

It is clearly evident from Fig.~\ref{bs_with_cp_for_different_N_without_blinking} that there is a sharp transition from a reasonably symmetric state (where $\langle f^+ \rangle$ is close to $0.5$) to a completely asymmetric state characterized by $\langle f^+ \rangle \sim 0$ as $\varepsilon_\text{ext}$ increases. This suggests that the external environment plays a key role in breaking the symmetry of the Oscillator Death state, as this phenomenon emerges only when the oscillator-environment coupling is sufficiently strong, with the sudden onset of asymmetry in the group of oscillators occurring at a critical coupling strength. 
Further, it is clear from Fig.~\ref{bs_with_cp_for_different_N_without_blinking} that the symmetry breaking of the Oscillator Death states is independent of system size $N$, over a large range of system sizes.
  
\section{Effect of Blinking Connections}

In the section above we considered the effect of the external environment on a group of oscillators, when the external system was connected to all oscillators at all times, and we clearly demonstrated that this lead to marked asymmetry in Oscillator Death states. This is in contradistinction to the case where the group of oscillators are not connected to an external system, which leads to complete symmetry in the Oscillator Death states. Now we will consider the  effect of oscillator-environment connections blinking on-off, and explore the effect of such time-varying links on the symmetry of the Oscillator Death states. 

In order to model connections to the environment blinking on and off, we consider a time-dependent oscillator-environment coupling strength $\varepsilon_\text{ext}^i (t) = \varepsilon_\text{ext} \ g_i (t)$ in Eqn.~\ref{eqn_system}, with the feedback from the oscillator group to the external system given by $\bar{y}=\frac{1}{N}\sum_{i=1}^N \varepsilon^i_\text{ext}(t) y_i$ . If the connection of an oscillator to the environment is constant, $g_i(t) = 1$ for all $t$. Such a link is considered a time-invariant {\em non-blinking connection}. If the connection of the $i^{\rm{th}}$ oscillator in the group and the external system periodically switches on and off, namely the link is a {\em blinking connection},  $g_i (t)$ is a square wave. When oscillator $i$ in the group is connected to environment $g_i (t)=1$, otherwise $g_i (t)=0$. So $g_i (t)$ switches between $0$ (off) and $1$ (on), with time period $T_\text{blink}$ which provides a measure of the time-scale at which the links vary. Here we will principally consider rapidly switching links, i.e. low $T_\text{blink}$. 

        
One of the most important parameters in this time-varying scenario is the fraction of blinking oscillator-environment connections in the group, which we denote by $f_\text{blink}$. If all oscillators are connected to the external, then $f_\text{blink}=0$ and if all connections are blinking, then $f_\text{blink}=1$. Here we will study the entire range $0 \le f_\text{blink} \le 1$, and gauge the effect of the fraction of blinking connections on the symmetry of OD state. Notice that the presence of connections switching on-off in a sub-set of oscillators results in the dynamical equations of the oscillator groups being less symmetric, as the group splits into two sub-sets having distinct dynamics. So it is most relevant to investigate if this lack of symmetry in the dynamical equations leads to more asymmetry in the steady states. However, what we will demonstrate in this Section is the following result: counter-intuitively, blinking links partially {\em restore } the symmetry of the emergent Oscillator Death states.



\begin{figure}[H]
    \centering
\includegraphics[width=1.0\linewidth]{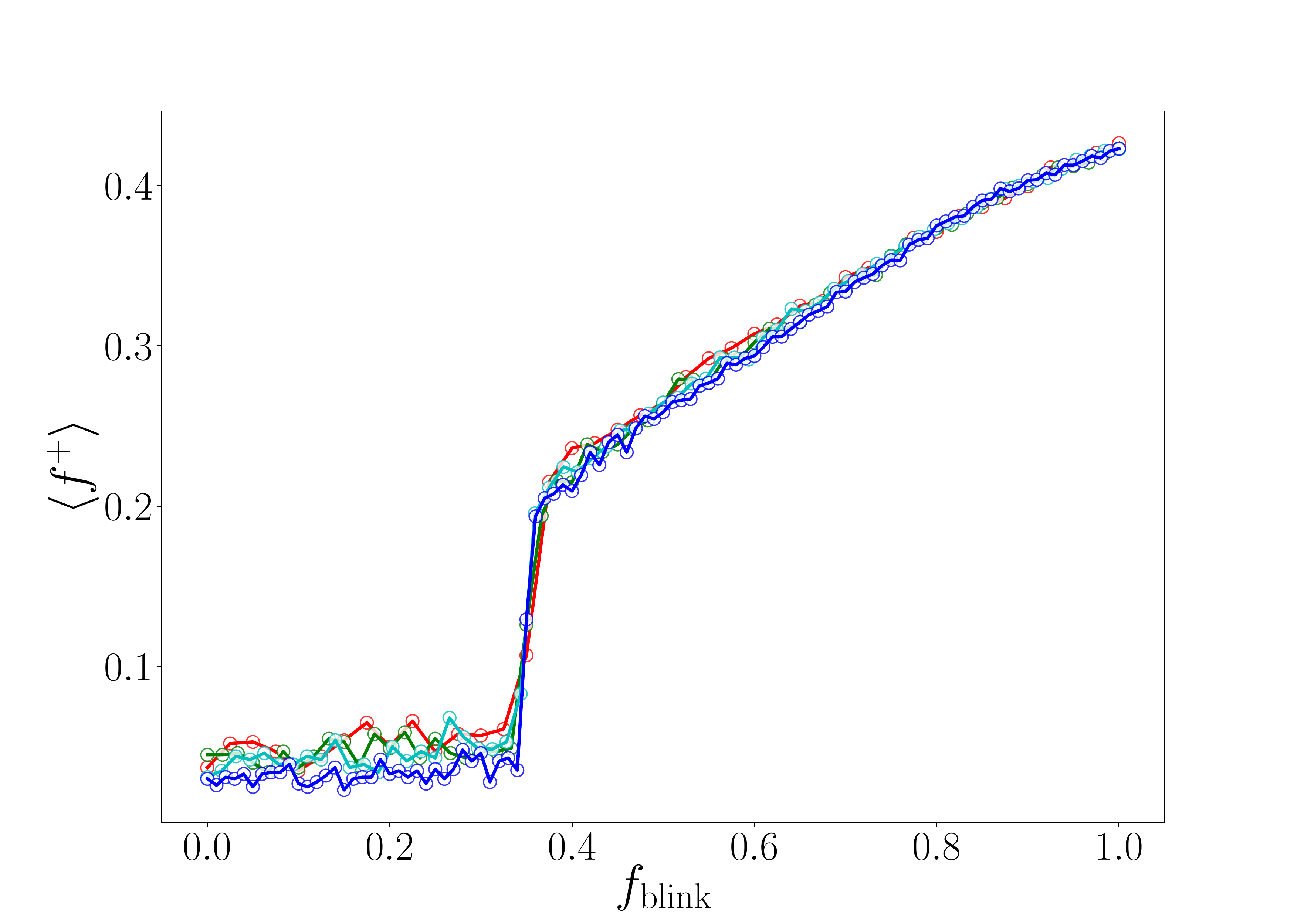}\\
\caption{Dependence of the Basin Stability of the positive Oscillator Death state $\langle f^+ \rangle$, on the fraction of oscillators $f_\text{blink}$ with blinking oscillator-environment connections in the group. 
Different system sizes $N=40,60,64,100$ are shown in different colours. Here the time period of the on-off blinking $T_\text{blink}=0.02$, oscillator-environment coupling strength $\varepsilon_\text{ext}=0.5$ and the damping constant of the environment $k=0.2$.}
    \label{bs_varying_blinkers_with_different_N}
  \end{figure}

Fig.~\ref{bs_varying_blinkers_with_different_N} shows the Basin Stability of the positive oscillator death state, as a function of the fraction $f_\text{blink}$ of oscillators with blinking connections to the environment. We find that when there are no blinking links, namely the connections of the oscillators to the external system are always on, the emergent state is the {\em most asymmetric}. That is, the deviations of the Basin Stability from $0.5$ is the most pronounced for $f_\text{blink} = 0$. Increasing the number of blinking connections reduces the asymmetry and restores the symmetry of the oscillator death states to a large extent, yielding states that are almost equi-distributed between positive and negative states. The transition from the asymmetric state (where $\langle f^+ \rangle \sim 0$) to a more symmetric state (where $\langle f^+ \rangle$ is significantly different from $0$) occurs sharply at a critical fraction of blinking links, which we denote by $f_\text{blink}^c$.
Further, it is evident from Fig.~\ref{bs_varying_blinkers_with_different_N} that the symmetry breaking dynamics of the system, and $f_\text{blink}^c$ in particular, is independent of number $N$ of oscillators in the group.

\begin{figure}[H]
    \centering
    \includegraphics[width=1.0\linewidth]{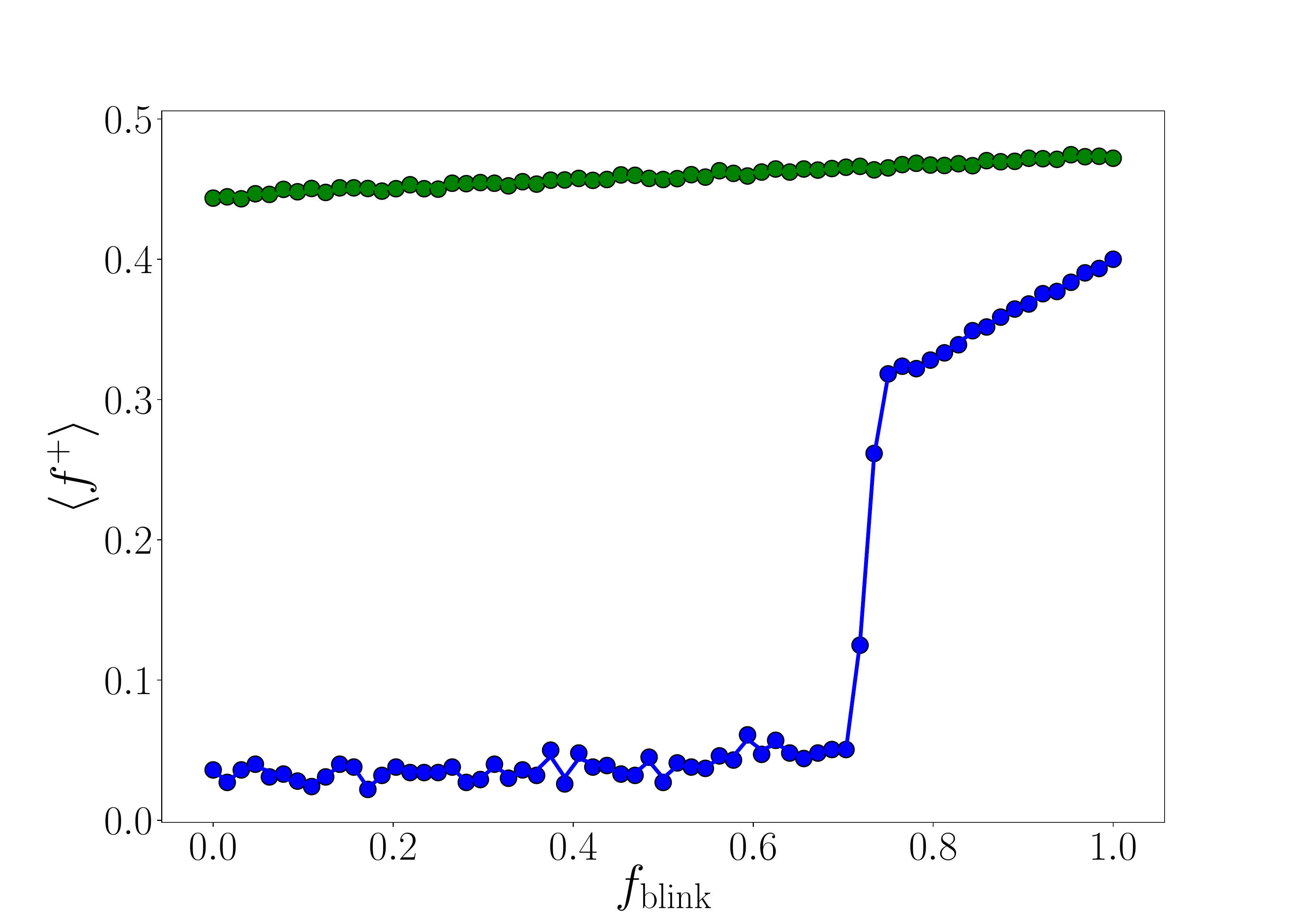}\\
    (a)\\
    \includegraphics[width=1.0\linewidth] {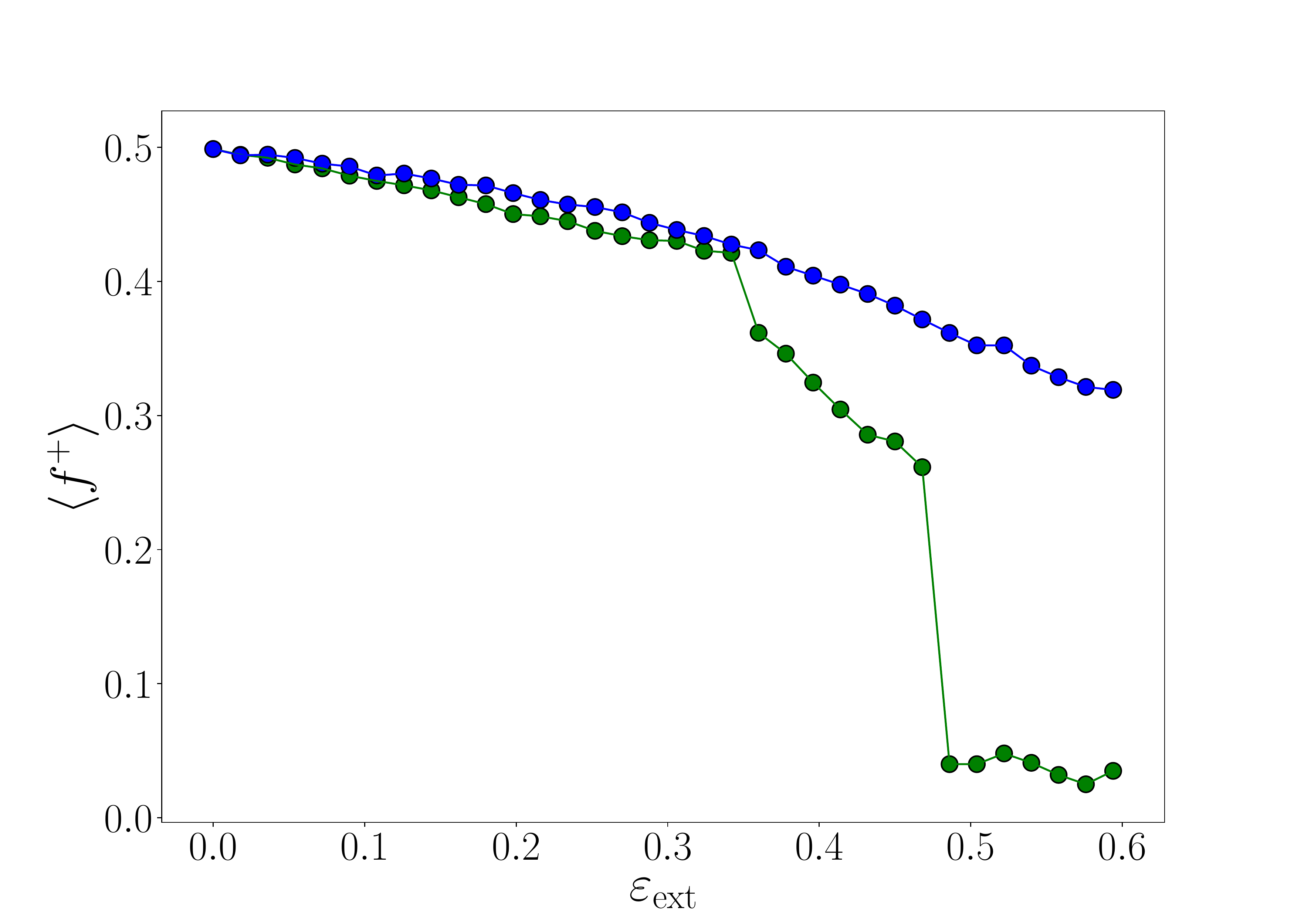}\\
    (b)
    \caption{Basin Stability of the positive Oscillator Death state, (a) as a function of fraction of oscillators with blinking oscillator-environment connections $f_\text{blink}$, for coupling strength $\varepsilon_\text{ext}=0.2$ (green) and 
$\varepsilon_\text{ext}=0.6$ (blue), and (b) as a function of  oscillator-environment coupling strength $\varepsilon_\text{ext}$, for $f_\text{blink}=0.25$ (green) and  $f_\text{blink}=0.75$ (blue). Here the time period of blinking $T_\text{blink}=0.02$, the 
damping constant of the environment $k=1.0$ and $N=64$. }
    \label{bs_varying_blinkers_with_different_E}
  \end{figure}

Fig.~\ref{bs_varying_blinkers_with_different_E} shows the dependence of the Basin Stability of the positive Oscillator Death state on the fraction of oscillators with blinking oscillator-environment connections $f_\text{blink}$, and the oscillator-environment coupling strength $\varepsilon_\text{ext}$. It is evident that for weaker coupling strengths the group of oscillators evolve to the positive and negative OD states with almost equal probability, i.e. $\langle f^+ \rangle$ is quite close to $0.5$. On the other hand, for strong coupling strengths, there is a sharp transition from a very asymmetric situation (where $\langle f^+ \rangle \sim 0$) at low $f_\text{blink}$, to a more balanced situation (where $\langle f^+ \rangle$ is closer to $0.5$) at high $f_\text{blink}$. Further, one observes that a system with a large fraction of blinking connections does not become markedly asymmetric even for large coupling strengths, i.e. $\langle f^+ \rangle$ is not close to $0$ even for $\varepsilon_\text{ext}$ close to $1$. However, in a system with few blinking connections, there is a sharp transition to asymmetry for sufficiently high coupling strengths.

We will now focus on the effect of the dynamical features of the common environment on symmetry breaking of the OD state. Note that  the common environment, when uncoupled to the oscillator group, is an exponentially decaying system $u=u_0 e^{-kt}$, where $u_0$ is the amplitude at time $t=0$ and $k$ is the damping rate. 

\begin{figure}[H]
    \centering
    \includegraphics[width=1.0\linewidth]
    {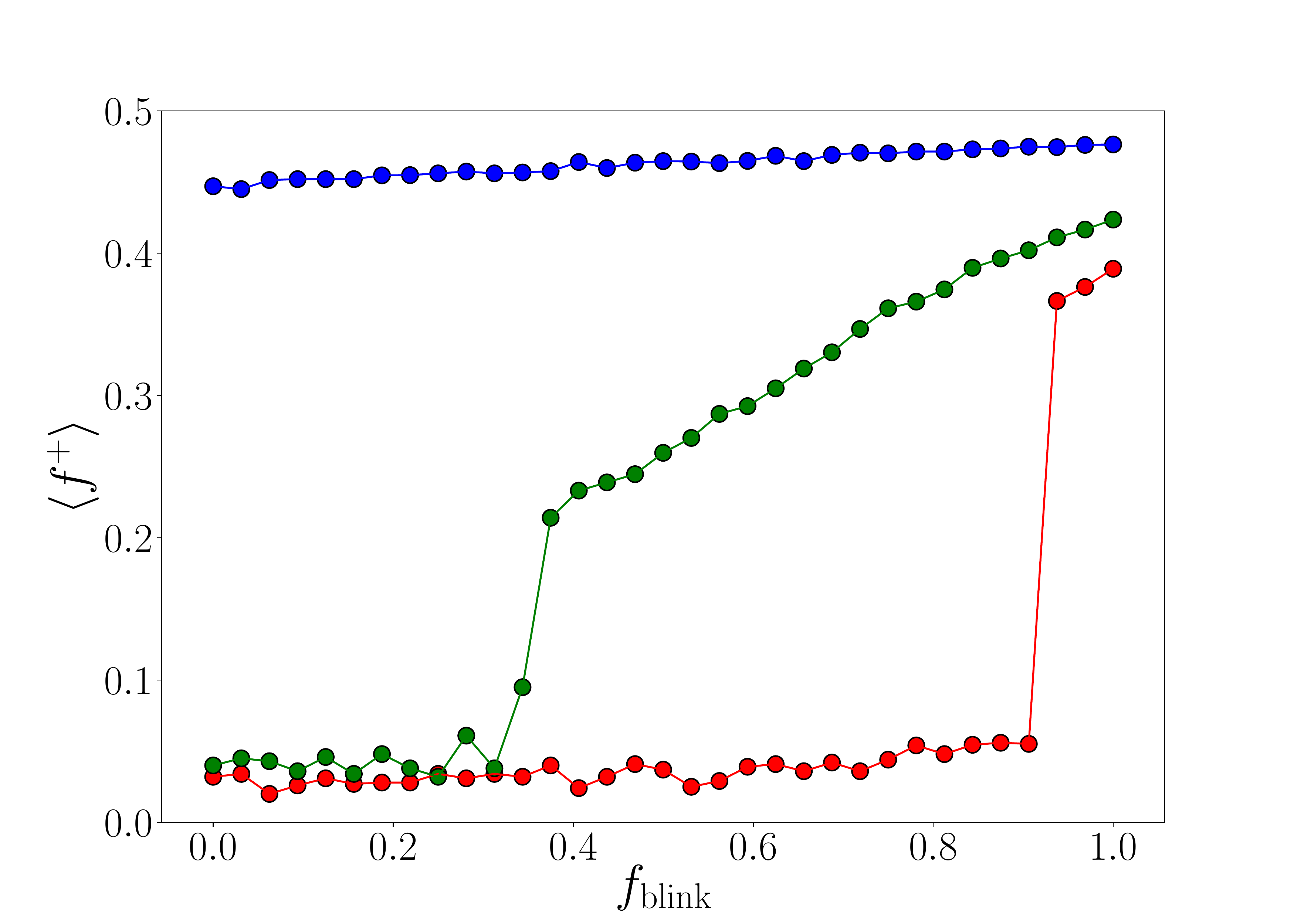}
    \caption{Basin Stability of the positive Oscillator Death state on the fraction of oscillators with blinking connections $f_\text{blink}$, for environmental damping constant $k = 0.1$ (red), $k = 0.2$ (green) and $k = 1.0$ (blue). The time period of blinking $T_\text{blink}=0.02$, oscillator-environment coupling strength  $\varepsilon_{ext} = 0.5$ and number of oscillators in the group $N = 64$. To estimate the Basin Stability we randomly sampled $u_0\in(0,1]$, for each $k$.}
\label{bs_varying_blinkers_with_different_K}
  \end{figure}

Fig.~\ref{bs_varying_blinkers_with_different_K} shows the probability of obtaining the positive Oscillator Death state, as the fraction of oscillators with blinking connections is varied, for different environmental damping constants $k$. It is evident that at high environmental damping rates, the effect of environment is less pronounced, and the Oscillator Death states are selected with almost equal probability. However, there is pronounced asymmetry in OD states when the damping rate of the environment is low, with critical  $f_\text{blink}^c$ tending to $1$ as $k$ increases.

  
  
  
  
\begin{figure}[H]
    \centering
    \includegraphics[width=1.0\linewidth] {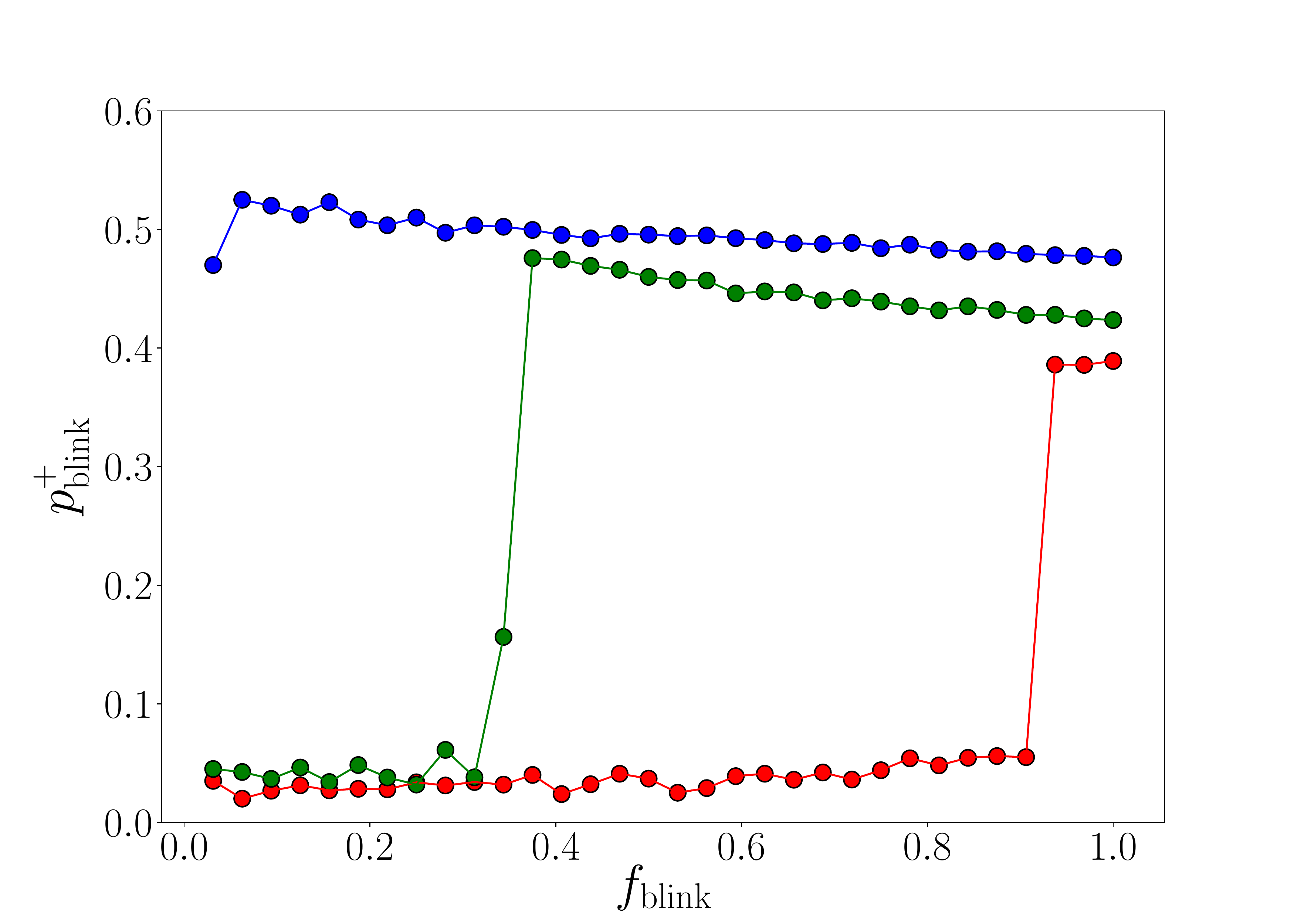}\\
    (a)\\
    \includegraphics[width=1.0\linewidth] {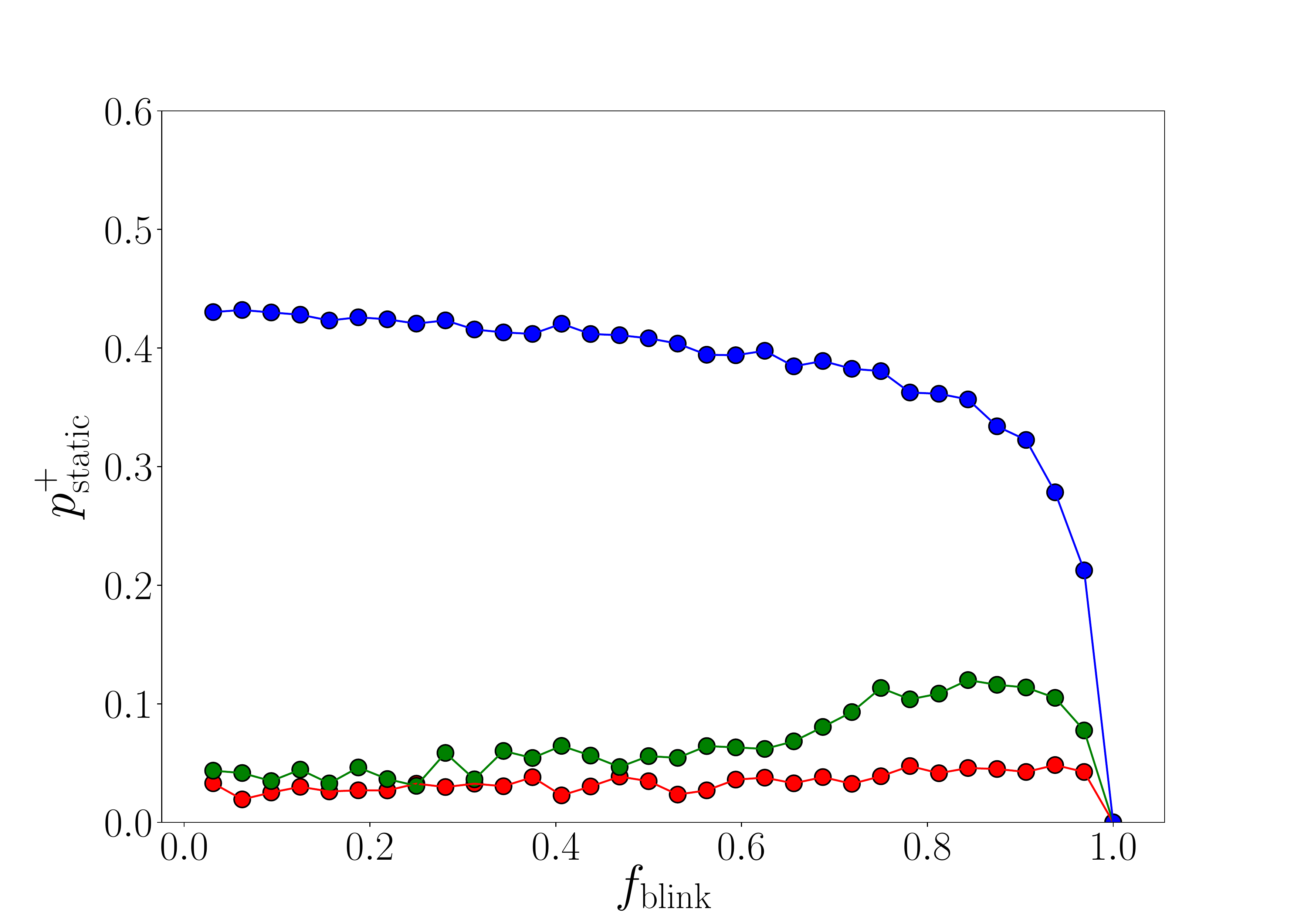}\\
    (b)
    \caption{Probability of obtaining the positive Oscillator Death state for the sub-group of oscillators with (a) blinking oscillator-environment connections, denoted by $p^+_{blink}$, and 
    (b) static oscillator-environment connections denoted by $p^+_{static}$, as a function of the fraction of oscillators with blinking connections $f_\text{blink}$. Here the time period of blinking $T_\text{blink}=0.02$, the oscillator-environment coupling strength $\varepsilon_{ext}=0.5$, number of oscillators in the group $N=64$, and  the environmental damping constant $k$=$0.1$ (red), $k=0.2$ (green) and $k=1.0$ (blue).}
    \label{probability_varying_blinkers_with_different_K}
  \end{figure}

Further we estimate the probability of an oscillator to be in the positive Oscillator Death state, for the case of the sub-group of oscillators with blinking links to the environment (see Fig.~\ref{probability_varying_blinkers_with_different_K}a), and for the case of the sub-group of oscillators with static links to the environment (see Fig.~\ref{probability_varying_blinkers_with_different_K}b).
  For low environmental damping rates, there is a sharp boost in the probability of oscillators to be in the positive OD state in the sub-group of oscillators with blinking oscillator-environment connections, at a critical $f_\text{blink}^c$ (e.g. $f_\text{blink}^c \sim 0.3$ for $k=0.2$ and $f_\text{blink}^c \sim 0.9$ for $k=0.1$.). For the case of the sub-group of oscillators with static oscillator-environment connections, the probability of obtaining the positive Oscillator Death state remains quite invariant. {\em This implies that the sub-group of oscillators with blinking connections to the environment is the group that is vital to the restoration of symmetry}.
  
    
\begin{figure}[H]
    \centering
    \includegraphics[width=1.0\linewidth]
    {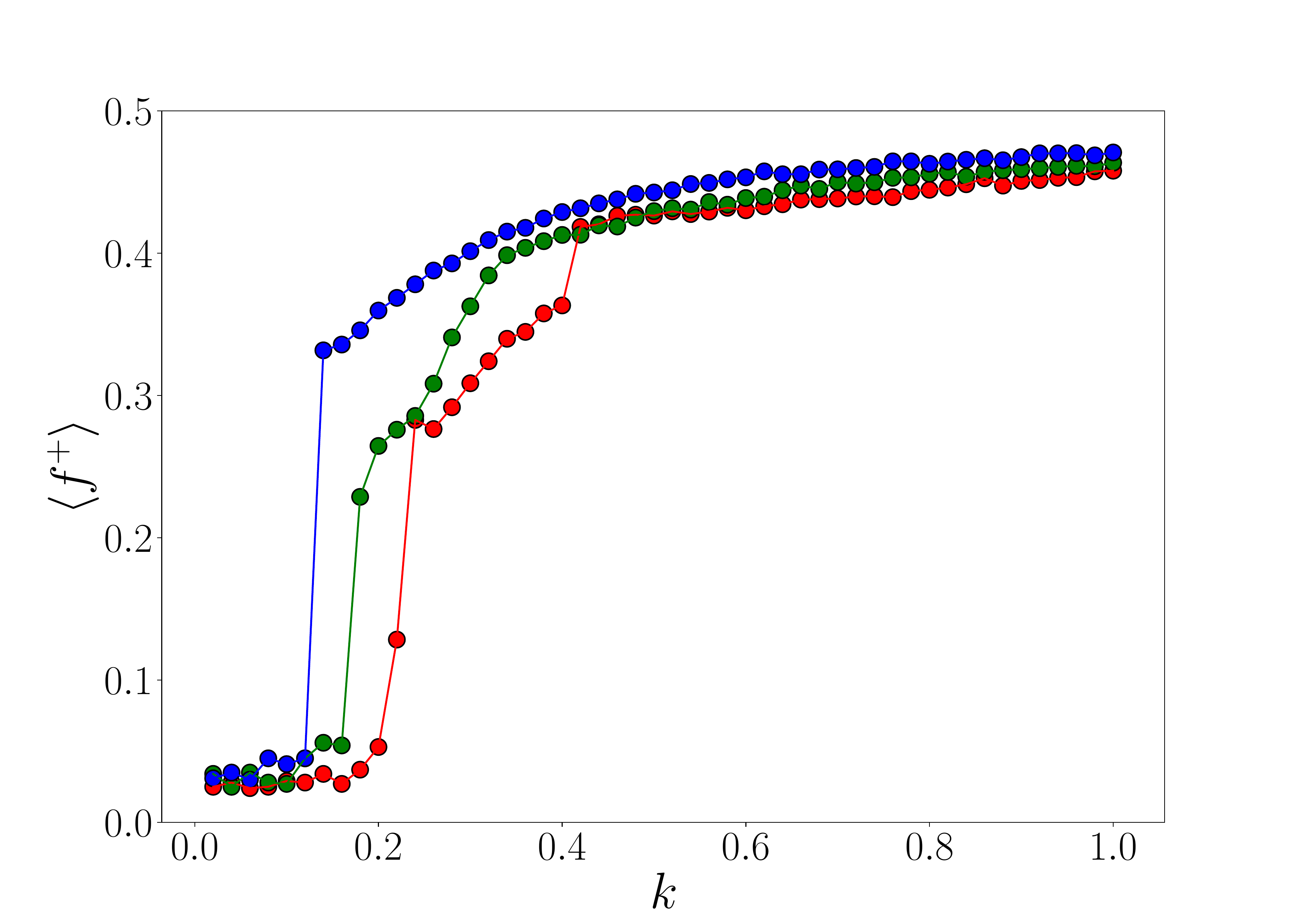}
    \caption{Basin Stability of the positive Oscillator Death state, as a function of the damping constant $k$ of the environment. Here time period of blinking $T_\text{blink}=0.02$, oscillator-environment coupling strength $\varepsilon_{ext} = 0.5$ and the number of oscillators in the group $N=64$ and the fraction of blinking oscillator-environment connections are: $f_\text{blink} = 0.25$ (red), $f_\text{blink} = 0.50$ (green) and $f_\text{blink} = 0.75$ (blue).}
    \label{bs_varying_K_with_different_Nblink}
  \end{figure}

Fig.~\ref{bs_varying_K_with_different_Nblink} shows the dependence of the fraction of oscillators in the positive Oscillator Death state on the damping constant $k$ of the environment, for different fractions of blinking connections. It is evident that at low damping constants, there are very few oscillators in the positive OD state. On increasing the damping constant of the environment there is a sharp jump in the fraction of  oscillators in the positive OD state. So there is a sudden transition from a very asymmetric state, where the fraction of oscillators in the positive OD state is close to zero, to a more symmetric state, where this fraction is close to half. The critical $k$ where this jump occurs depends upon the number of blinking oscillator-environment links. When there is a higher fraction $f_\text{blink}$ of blinking connections in the system, the critical $k$ is lower, i.e. the jump to a more symmetric situation occurs at lower damping constants.
 
    \begin{figure}[H]
    \centering
    \includegraphics[width=1.0\linewidth]
    {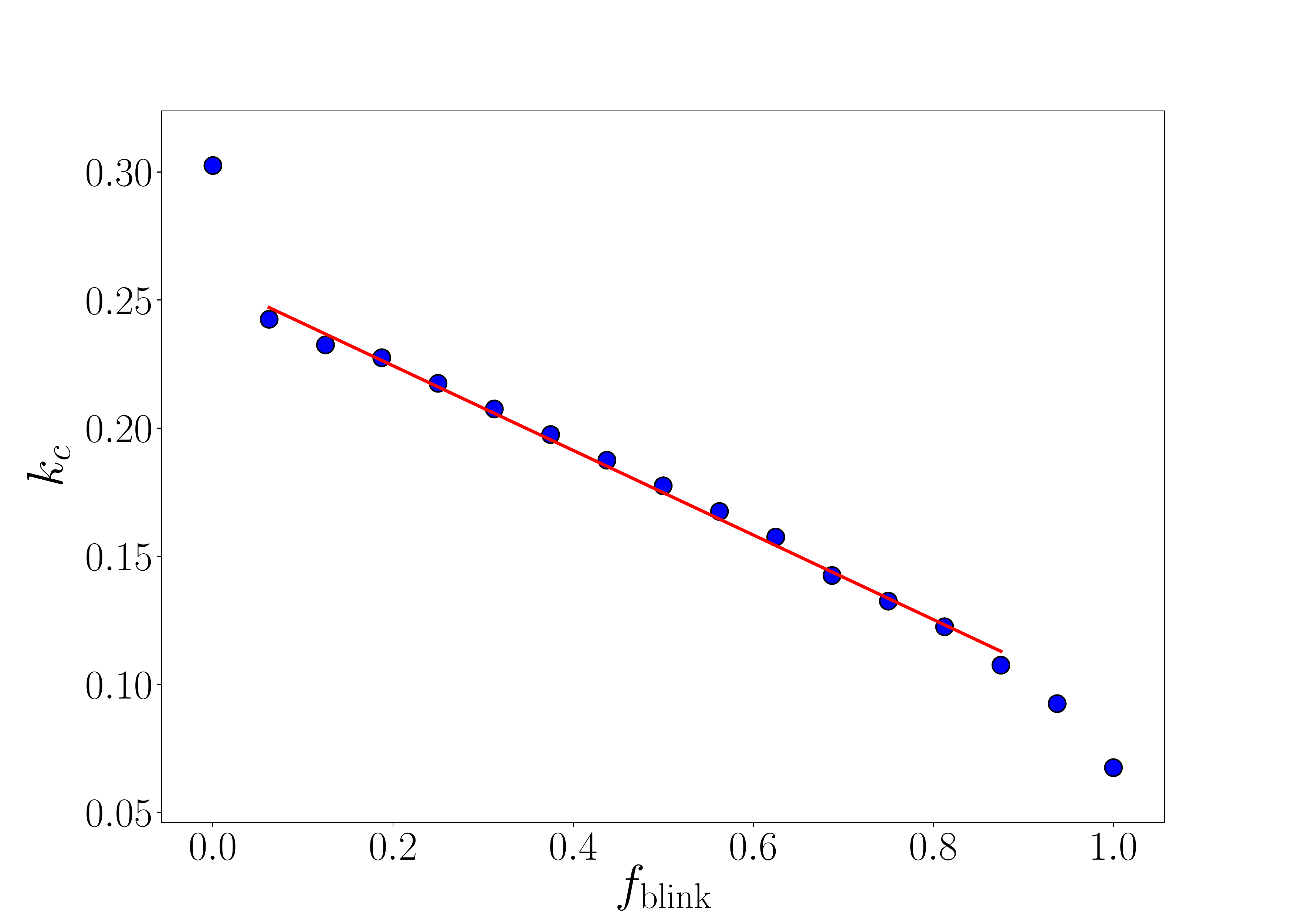}
    \caption{Critical value of damping constant $k_c$ vs. fraction of blinking oscillator-environment connections $f_\text{blink}$. Here the time period of blinking $T_\text{blink}=0.02$, oscillator-environment coupling strength $\varepsilon_{ext} = 0.5$ and number of oscillators in the group $N = 64$. The data points from numerical simulations are in blue, and the curve given by equation: $k_c= k_c^0 - c \ f_\text{blink}$, for $k_c^0 \approx 0.26$ and $c \approx 0.16$, is shown in red.}
 \label{critical_damping_constant_for_sudden_transition_with_fblink}
  \end{figure}
  
Further we estimate the value of damping constant $k$ where $\langle f^+ \rangle$ crosses a threshold value of $0.1$ (with no loss of generality), denoted by $k_c$. This critical value indicates the damping constant below which significant symmetry-breaking of the Oscillator Death states occurs. Fig.~\ref{critical_damping_constant_for_sudden_transition_with_fblink} shows critical $k_c$ as a function of the fraction of blinking connections $f_\text{blink}$. The critical damping $k_c$ decreases with increasing fraction of blinking connections $f_\text{blink}$. Specifically, in a large range of $f_\text{blink}$ we find that $k_c$ decreases linearly with $f_\text{blink}$
(see Fig.~\ref{critical_damping_constant_for_sudden_transition_with_fblink}).
This demonstrates that low environmental damping favours enhanced asymmetry, while more blinking connections tends to restore the symmetry of the OD states.
  
  

\begin{figure}[H]
    \centering
    \includegraphics[width=1.0\linewidth]
    {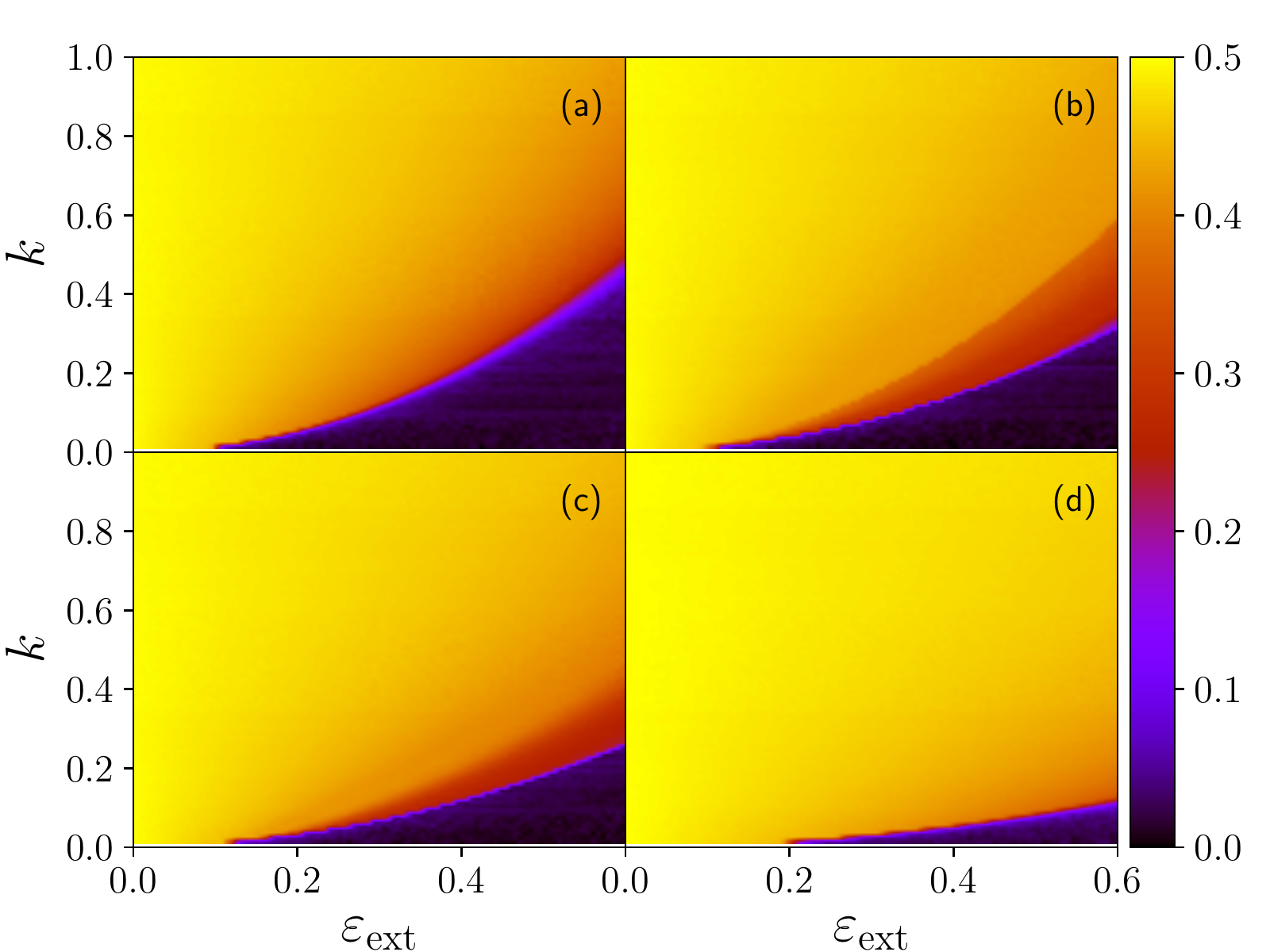}
    \caption{Basin Stability of the positive Oscillator Death in the parameter space of oscillator-environment coupling strength $\varepsilon_\text{ext}$ and environmental damping constant $k$. The fraction of blinking oscillator-environment connections are: (a) $f_\text{blink}=0.0$ (i.e. the static case), 
    (b) $f_\text{blink}=0.25$, (c) $f_\text{blink}=0.50$, (d) $f_\text{blink}=1.0$. The  time period of blinking $T_\text{blink}=0.02$ and the number of oscillators in the group $N=64$.}
    \label{bs_in_parameter_space_of_cp_and_k_with_different_fblink}
  \end{figure}

  
Fig.~\ref{bs_in_parameter_space_of_cp_and_k_with_different_fblink} shows the fraction of oscillators in the positive Oscillator Death state, in the parameter space of $k$ and $\varepsilon_\text{ext}$, for different fraction of blinking oscillator-environment connections. The black regions in the figures represent the asymmetric state. Clearly, low environmental damping $k$ and high oscillator-environment coupling $\varepsilon_\text{ext}$ yields the greatest asymmetry in the emergent Oscillator Death states. 
  
\begin{figure}[H]
    \centering
    \includegraphics[width=1.0\linewidth]
    {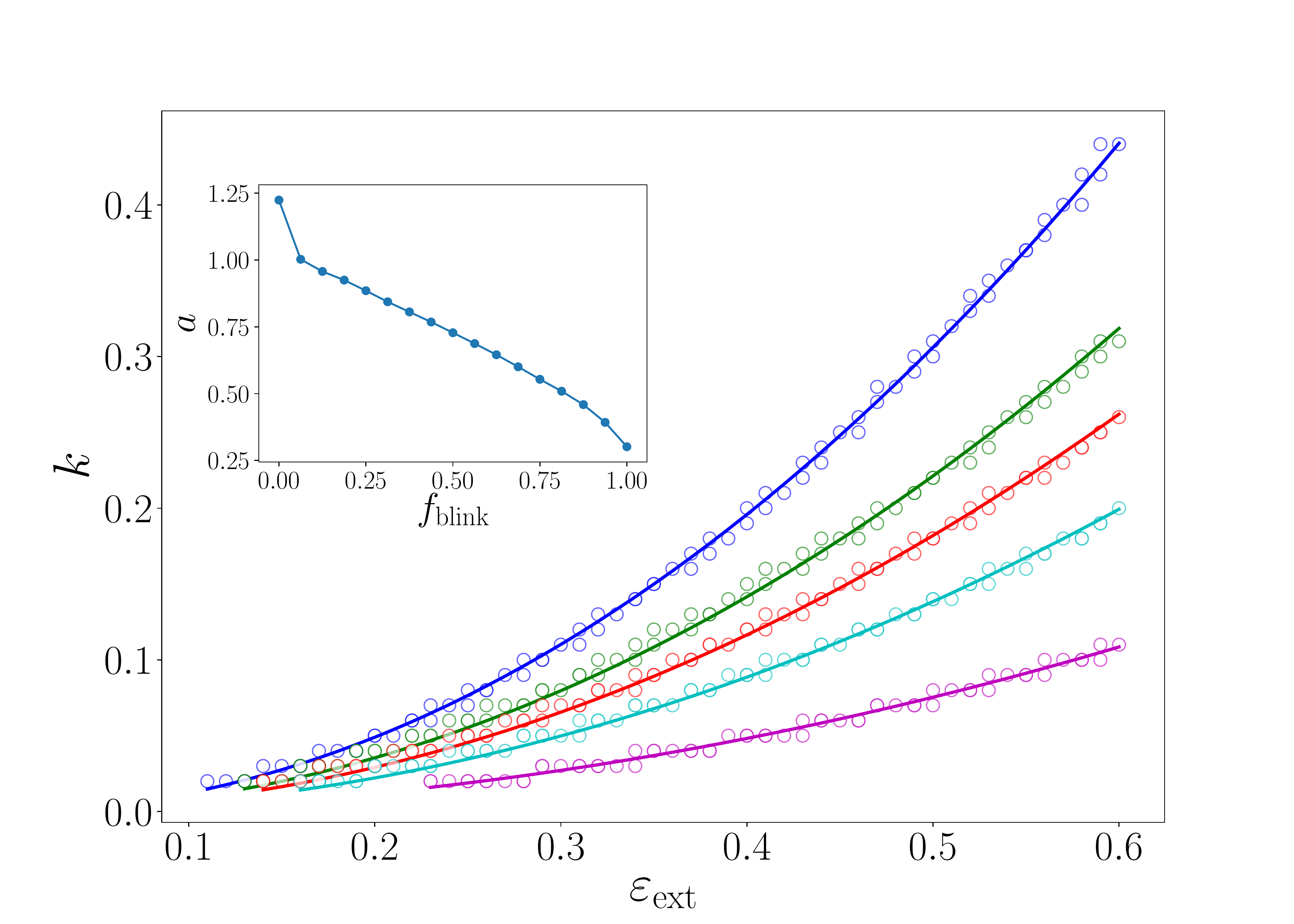}
    \caption{Transition line in Fig.~\ref{bs_in_parameter_space_of_cp_and_k_with_different_fblink}, fitted to the curve (solid lines): $k = a \varepsilon_\text{ext}^2$, where the different curves correspond to $f_\text{blink}=0.0,0.25,0.50,0.75,1.0$ from top to bottom. The inset shows the variation of $a$ with $f_\text{blink}$.}
    \label{transition line_in_bs_in_parameter_space_of_cp_and_k_with_different_fblink}
  \end{figure}

Now we focus on the line of transition from high $\langle f^+ \rangle$ to low $\langle f^+ \rangle$, shown in Fig.~\ref{transition line_in_bs_in_parameter_space_of_cp_and_k_with_different_fblink}. We find that $k$ is proportional to $\varepsilon_\text{ext}^2$ along the lines of transition, with the proportionality constant depending on the fraction of blinking oscillator-environment connections (see inset). Interestingly, comparing Fig.~\ref{critical_damping_constant_for_sudden_transition_with_fblink} and the inset of Fig.~\ref{transition line_in_bs_in_parameter_space_of_cp_and_k_with_different_fblink}, reveals that both have the same dependence on $f_\text{blink}$, and the proportionality constant is equal to $4k_c$.
This implies that the line of transition to asymmetry in the space of $k$-$\varepsilon_\text{ext}$ is given by: 
   \begin{equation}
   \label{k_and_kc_relation}
   k= 4 k_c  \ \varepsilon_\text{ext}^2
   \end{equation}
where $k_c$ is inversely proportional to the fraction of blinking connections $f_\text{blink}$.

\begin{figure}[H]
    \centering
    \includegraphics[width=1.0\linewidth]
{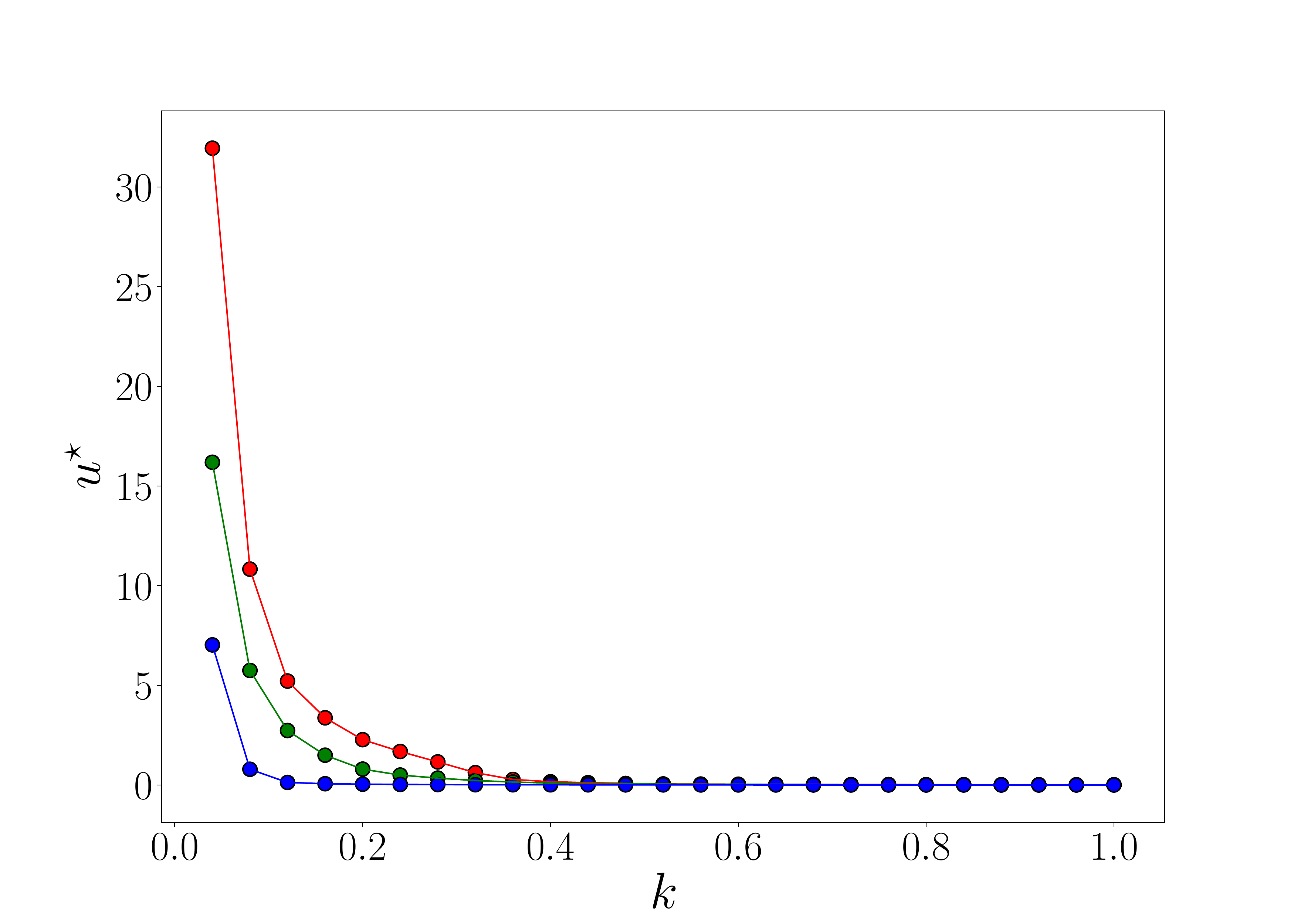}\\(a)\\
	\includegraphics[width=1.0\linewidth]{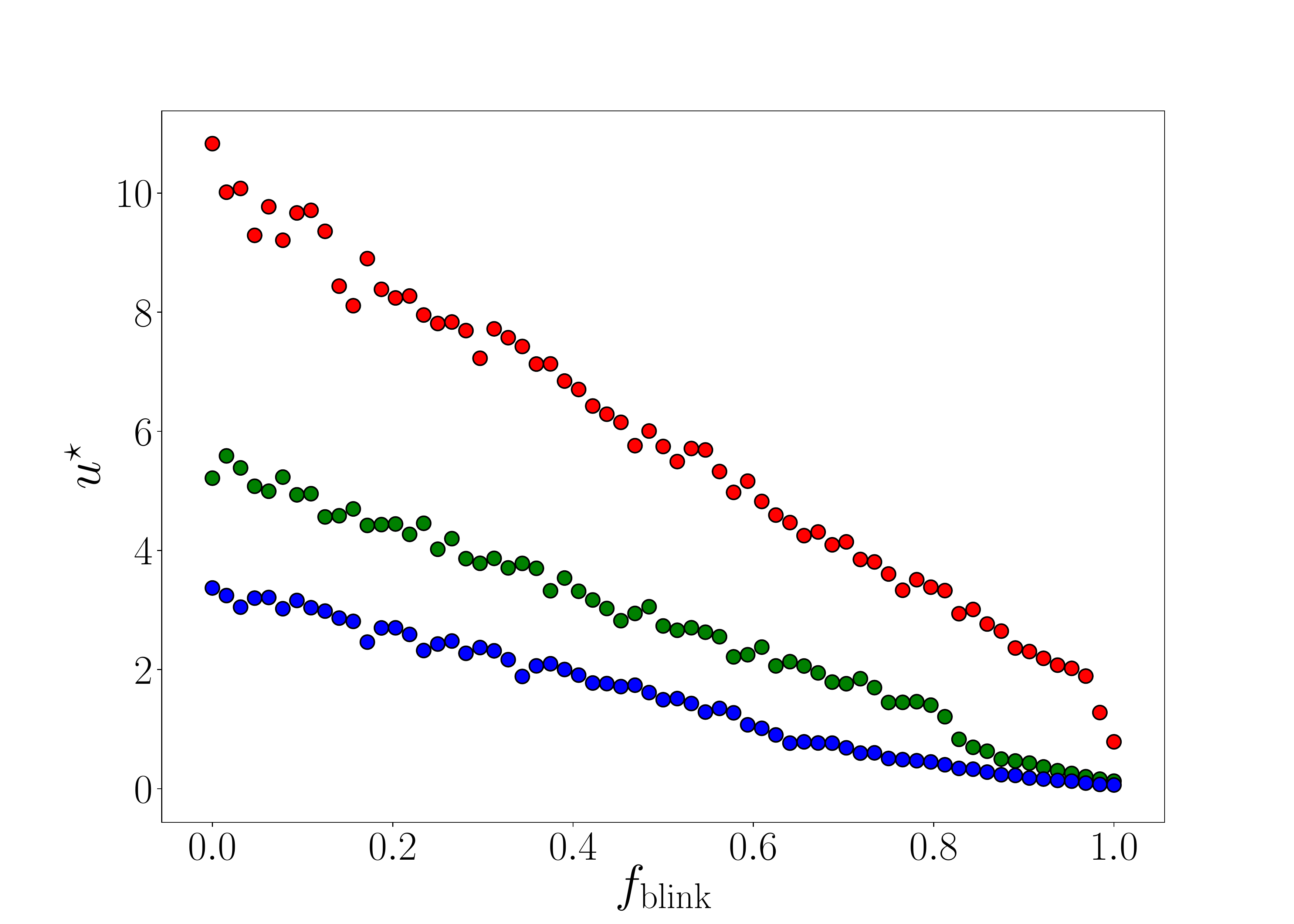}\\(b)
    \caption{Environmental steady state $u^{\star}$ with respect to (a) damping constant $k$, for $f_{blink}=0.0$ (red), $0.5$(green), $1.0$(blue) and (b) with respect to the fraction of oscillator-environment blinking connections, for damping constant $k=0.08$ (red), $0.12$ (green), $0.16$ (blue). Here the number of oscillators in the group $N=64$.
   }
    \label{offset_varying_k_and_blinkers}
  \end{figure}

Lastly, we explore the effect of this symmetry-breaking on the external environment. The external system is a damped system influenced by the mean field of the group of oscillators, and it settles down to a fixed point $u^{\star}$ when the OD state becomes stable. This is easily seen as follows: when the OD state is stable, $\bar{y}$ is a constant. So the steady state solution of $u$ is given by $\bar{y}/k$. Denoting the $x$ variable of the positive OD state by $x^+$ and the $y$-variable as $y^+$, and denoting the $x$ variable of the negative OD state by $x^-$ and the $y$-variable as $y^-$, we have $\bar{y} = f^+ y^+ + (1-f^+)y^-$, yielding $u^{\star} = \frac{y^+ (2 f^+-1)}{k}$. Clearly if the probabilities of obtaining the positive and negative OD states are equal, i.e. $f^+=0.5$, then $u^{\star}=0$. If the asymmetry is extreme and $f^+ \sim 0$ we have $u^{\star} =-y^+/k$. Since the positive OD state has $x^+ >0$ and $y^+ < 0$, $u^{\star}$ is positive. Also notice that the value of $u^{\star}$ is inversely proportional to damping constant $k$ (cf. Fig.~\ref{offset_varying_k_and_blinkers}a). So a strongly damped environment evolves to $u^{\star}$ close to zero, as is intuitive. Further, from linear stability analysis of the dynamics of the external system one can see that $u^{\star}$ is a stable steady state, as the derivative if the vector field governing $\dot{u}$ is $-k$ which is always negative. Further, since increasing the fraction of blinking connections favours symmetry, thereby pushing $f^+$ closer to half, $u^{\star}$ also decreases with increasing $f_\text{blink}$ (cf. Fig. \ref{offset_varying_k_and_blinkers}b).

So we can conclude that the state of the environment $u^{\star}$ is {\em strongly correlated to the asymmetry}. In fact, simply observing $u^{\star}$ tells us if the symmetry-breaking is pronounced or not.

\bigskip

\noindent
{\em Effect of the frequency of blinking on Oscillation Death:}\\

\begin{figure}[H]
    \centering
\includegraphics[width=1\linewidth] {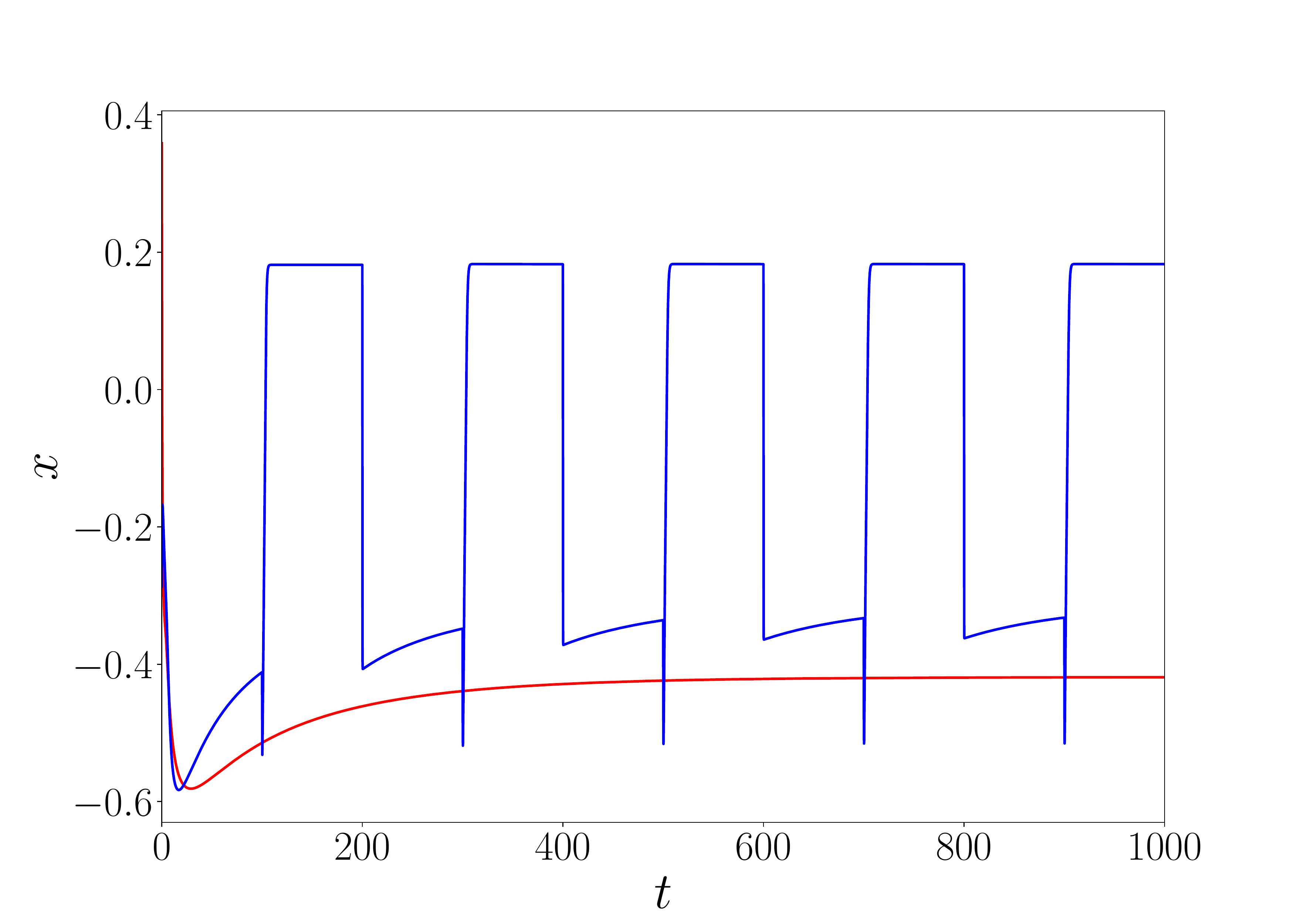}\\ 
     \hfill (a) \hfill $ $\\
    \includegraphics[width=0.95\linewidth] {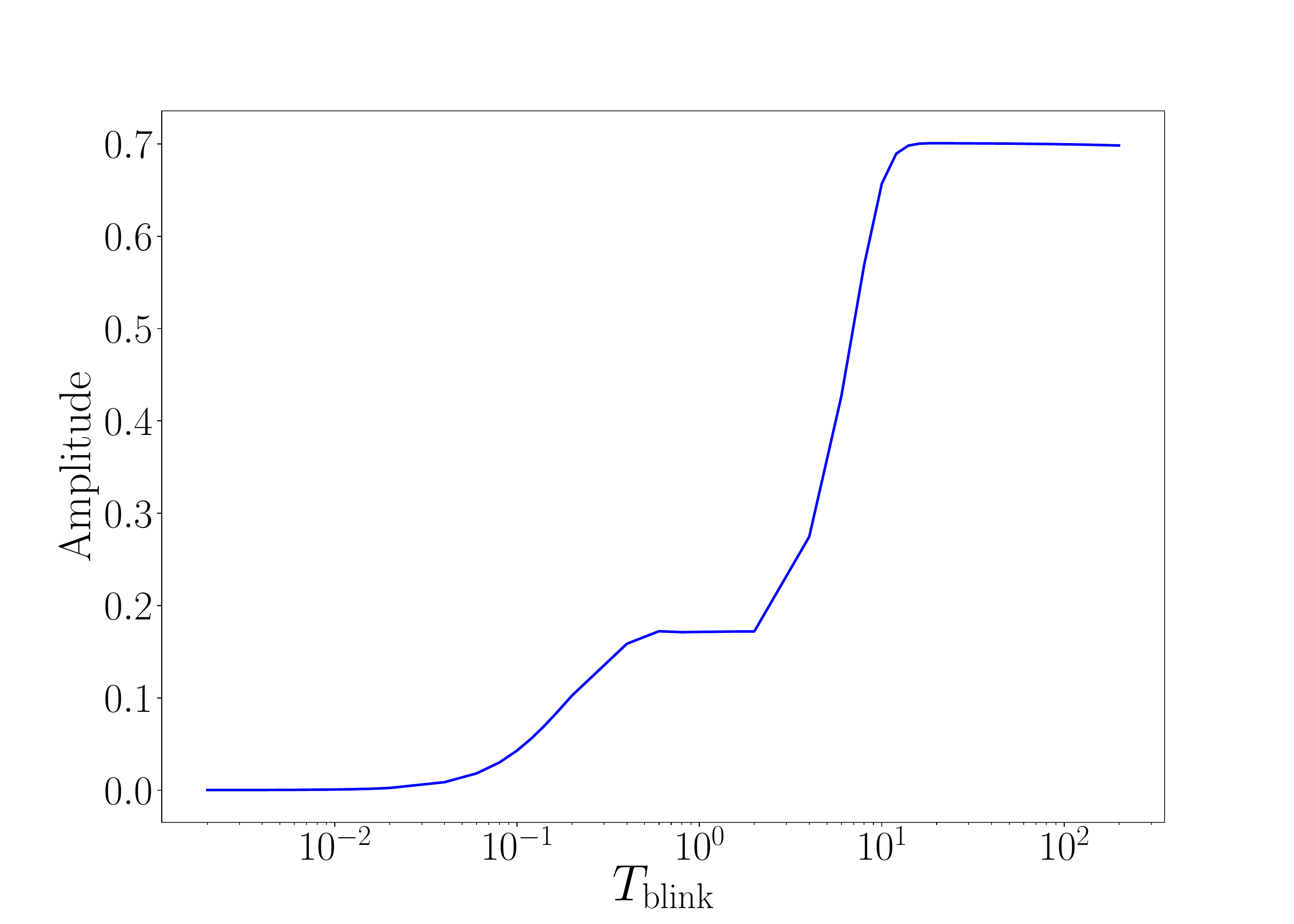}\\
    \hfill (b) \hfill $ $
	\caption{(a) Time series of one of the oscillators in group for $T_\text{blink}=0.02$ (red), $200.0$ (blue) and (b) Amplitude of the $x$-variable for different $T_\text{blink}$. Here 
$N=20$, $f_\text{blink}=0.1$, $\varepsilon_\text{ext}=0.6$ and $k=0.01$.}
    \label{effect_of_switching_time}
\end{figure}

In the results discussed above the time-period of the blinking connections was small, i.e. the links switched on-off rapidly. Now we will investigate the influence of the time-period $T_\text{blink}$ of the blinking oscillator-environment connections on the dynamics. Fig.~\ref{effect_of_switching_time} displays the effect of increasing blinking time-period on the state of the oscillators. It is evident from the time-series of the oscillators (cf. Fig.~\ref{effect_of_switching_time}a) that after a critical blinking time-period the system starts to oscillate and the OD steady state is destroyed. That is, {\em slow blinking of links leads to oscillation revival.} This is also quantitatively demonstrated in Fig.~\ref{effect_of_switching_time}b, which shows the amplitude of the oscillators. Clearly up to $T_\text{blink} \sim 0.1$ the amplitude is zero and one obtains a Oscillator Death state. However, when $T_\text{blink}$ increases further, the amplitude grows from zero to a finite value, indicating the emergence of oscillations whose amplitude increases with $T_\text{blink}$. After a large value $T_\text{blink}$ ($\sim 10$) the amplitude of the oscillations saturate to a maximum value (cf. Fig.~\ref{effect_of_switching_time}b). We find that this maximum amplitude is the difference between the steady state solution of the oscillator for the case of $\varepsilon_\text{ext} = 0$ (i.e. when uncoupled from the external system) and the steady state arising for $\varepsilon_\text{ext} > 0$ (i.e. when the oscillator group is coupled to a common environment). So the oscillator moves periodically between the two steady state solutions when the blinking is slow enough to allow the system to reach the two distinct steady states during the on and off period respectively.


Further, notice that there are two distinct transitions to oscillation revival. The first occurs around $T_\text{blink} \sim 0.1$, with fixed states transitioning to non-zero amplitude oscillations, saturating around amplitude $\sim 0.2$. The second transition commences around $T_\text{blink} \sim 2 \pi/\omega$, where the amplitude starts to grow rapidly again, from amplitudes around $0.2$, to the maximum amplitude.

\section{Constant Common environment}
  
We have shown the effect of exponentially decaying external environment on the OD state and the effect of blinking connections. This mimics a situation where the external environment is a small bath, and so the dynamics of the oscillator group affects the dynamics of the common environment. In this section, we will consider a common external environmental system mimicking a large bath, where the external environment does not get affected by the oscillator group. Rather it acts as a {\em constant drive}, which we denote by $u_c$. The strength of this oscillator-external drive connection is given by the coupling strength $\varepsilon_{\text{ext}}$. So the complete dynamics of the group of oscillators is now given by the following evolution equations:

    \begin{eqnarray}
        \label{eqn_constant_drive}
        \dot{x}_i &=& (1-x_i^2-y_i^2)x_i-\omega y_i + \varepsilon_\text{intra}(q\bar{x}-x_i) \nonumber \\
        \dot{y}_i &=& (1-x_i^2-y_i^2)y_i+\omega x_i + \varepsilon_\text{ext} u_c
    \end{eqnarray}

    where $\bar{x}=\frac{1}{N}\sum_{i=1}^N x_i$.

      \begin{figure}[H]
    \centering
    \includegraphics[width=1.0\linewidth]
    {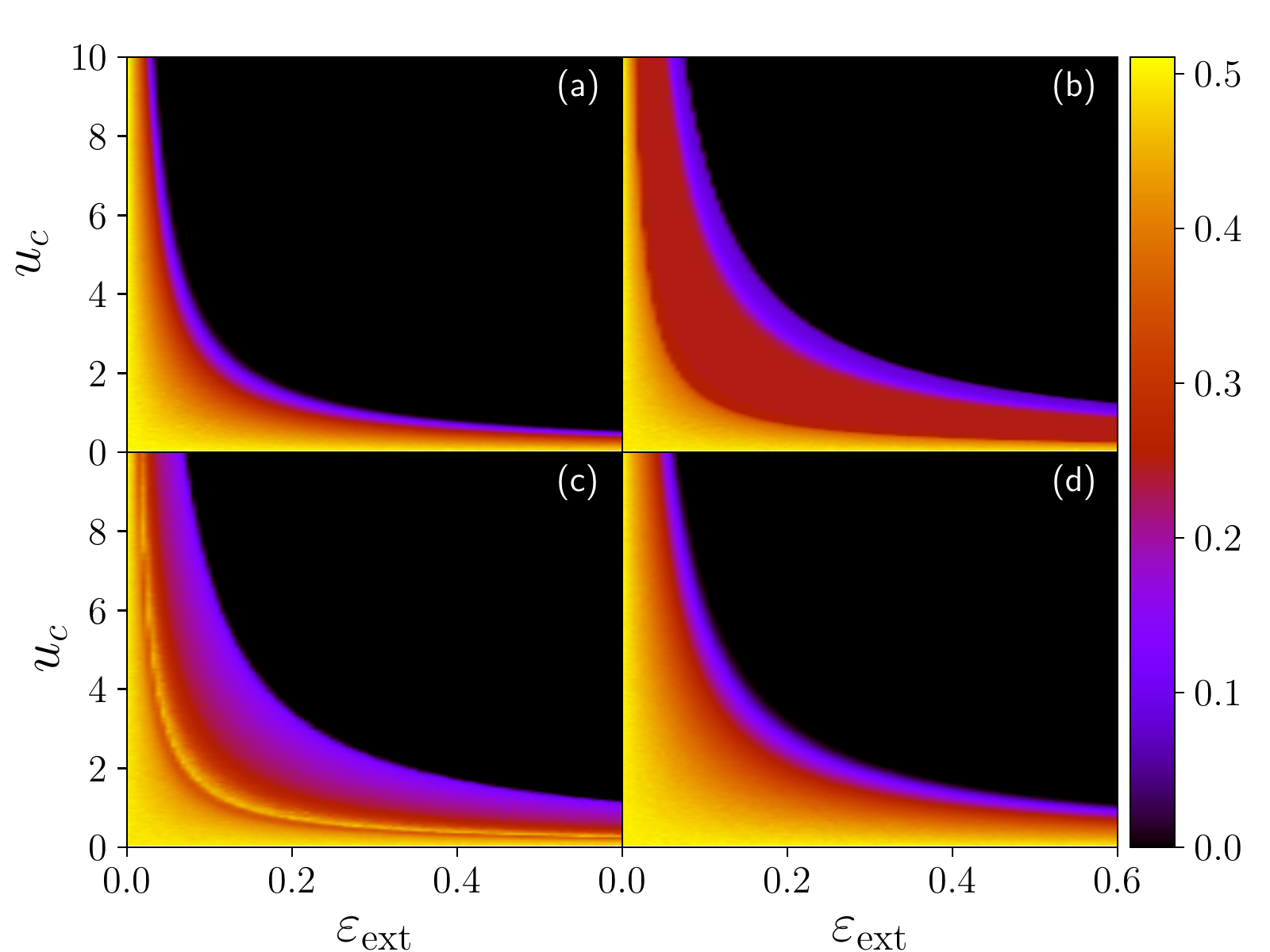}
    \caption{Basin Stability of the positive Oscillator Death state in the 
    parameter space of coupling strength $\varepsilon_\text{ext}$ and constant 
    environment ($u_c$), with fraction of blinking oscillators (a) $f_\text{blink}=0.0$, 
    (b) $f_\text{blink}=0.25$, (c) $f_\text{blink}=0.50$, (d) $f_\text{blink}=1.0$. Here the time period of blinking $T$=$0.02$ and the number of oscillators in the group $N=64$.}
    \label{bs_in_parameter_space_of_cp_and_uc_with_different_fblink}
  \end{figure}
  
Fig.~\ref{bs_in_parameter_space_of_cp_and_uc_with_different_fblink} shows the fraction of oscillators in positive Oscillator Death state, in the parameter space of $u_c$ and $\varepsilon_\text{ext}$. Different panels correspond to different fraction of blinking connections, and the black regions in the figures represent the asymmetric state. It is evident that even for the case of constant drive, the symmetry of the Oscillator Death state is broken, and the system moves preferentially to the negative state. This further indicates the generality of our observations, and emergence of symmetry-breaking in a group of oscillators due to coupling to a common external system.


\bigskip

\section{Conclusion}

We investigated the impact of a common external system, which we call a {\em common environment},  on the Oscillator Death states of a group of Stuart-Landau oscillators. First we consider external systems that exponentially decay to zero when uncoupled from the oscillator group. Note that a group of oscillators yield a completely symmetric Oscillator Death state when uncoupled to the external system, i.e. the positive and negative OD states occur with equal probability, and so in a large ensemble of oscillators the fraction of oscillators attracted to the positive/negative state is very close to half. However, remarkably, when coupled to a common external system this symmetry is significantly broken. This symmetry breaking is very pronounced for low environmental damping and strong oscillator-environment coupling, as evident from the sharp transition from the symmetric to asymmetric state occurring at a critical oscillator-environment coupling strength and environmental damping rate.
  
Further, we consider a group of oscillators with time-varying connections to the common external environment. In particular, we study the system with a fraction of oscillator-environment links that switch on-off. Interestingly, we noticed that the asymmetry induced by environmental coupling decreases as a power law with increase in fraction of such on-off connections. This suggests that blinking oscillator-environment links can restore the symmetry of the Oscillator Death state.

Lastly, we demonstrated the generality of our results for a constant external drive, i.e. a constant environment, and found marked breaking of symmetry Oscillator Death states there as well. When the constant drive is large, the asymmetry in OD-state is very large, and the transition between the  symmetric and asymmetric state, with increasing oscillator-environment coupling, is sharp.

In summary, we have shown the existence of a pronounced breaking of symmetry in the Oscillator Death states of a group of oscillators induced by a common external environment. So our results demonstrate an environment-mediated mechanism for the prevalence of certain states in a system of oscillators, and suggests an underlying process for obtaining certain states preferentially in ensembles of oscillators.

\end{document}